**Momentum space separation of quantum path interferences between photons and surface plasmon polaritons in nonlinear photoemission microscopy**

Pascal Dreher[1*], David Janoschka[1], Harald Giessen[2], Ralf Schützhold[1,4,5], Timothy J. Davis[1,2,3], Michael Horn-von Hoegen[1], Frank-J. Meyer zu Heringdorf[1,6]

[1] Faculty of Physics and Center for Nanointegration, Duisburg-Essen (CENIDE), University of Duisburg-Essen, 47048 Duisburg, Germany

[2] 4th Physics Institute, Research Center SCoPE, and Integrated Quantum Science and Technology Center, University of Stuttgart, 70569 Stuttgart, Germany

[3] School of Physics, University of Melbourne, Parkville, Victoria 3010, Australia

[4] Helmholtz-Zentrum Dresden-Rossendorf, Bautzner Landstrasse 400, 01328 Dresden, Germany

[5] Institut für Theoretische Physik, Technische Universität Dresden, 01062 Dresden, Germany

[6] Interdisciplinary Center for the Analytics on the Nanoscale (ICAN), 47057 Duisburg, Germany

[*] Corresponding Author. E-Mail: pascal.dreher@uni-due.de

**Abstract**

Quantum path interferences occur whenever multiple equivalent and coherent transitions result in a common final state. Such interferences strongly modify the probability of a particle to be found in that final state, a key concept of quantum coherent control. When multiple nonlinear and energy-degenerate transitions occur in a system, the multitude of possible quantum path interferences is hard to disentangle experimentally. Here, we analyze quantum path interferences during the nonlinear emission of electrons from hybrid plasmonic and photonic fields using time-resolved photoemission electron microscopy. We experimentally distinguish quantum path interferences by exploiting the momentum difference between photons and plasmons and through balancing the relative contributions of their respective fields. Our work provides a fundamental understanding of the nonlinear photon-plasmon-electron interaction. Distinguishing emission processes in momentum space, as introduced here, will ultimately allow nano-optical quantum-correlations to be studied without destroying the quantum path interferences.

**Introduction**

Quantum path interference occurs when multiple coherent pathways can take a system from an initial state to a final state, as epitomized in Feynman's path-integral interpretation [1]. Brumer and Shapiro realized in the last century that light can be used to create quantum path interferences in matter [2, 3]. Coherent control of such quantum path interferences has been demonstrated in photoexcitation experiments [2, 4-13] and, more recently in nonlinear photoemission [14-19]. In all cases, quantum path interference relies on not knowing which path a system takes. Measurements aiming at obtaining this information destroy the interference, as observed in "which-way" experiments [20].

In nonlinear electron emission [21-25] the simultaneous absorption of energy-degenerate photons and surface plasmon polaritons (SPPs) can create multiple quantum pathways [26-28]. As these pathways result in a common final state, quantum path interferences are expected to appear in the electron yield. For strong electromagnetic fields, the quantum path interferences could be derived by treating the electrons quantum mechanically while approximating the photon and SPP fields as classical fields. However, for a consistent picture these fields must be treated as quantum fields.

The multitude of possible nonlinear mixings of these different pathways also makes it difficult to experimentally disentangle the resulting quantum path interferences. Here, in analogy to optical nonlinear spectroscopy [29, 30], we use in a novel momentum space approach to separate such quantum path interferences in nonlinear photoemission electron microscopy (PEEM).

**Experimental Details**

The experiments were performed in a spectroscopic photoemission and low energy electron microscope (ELMITEC SPE-LEEM III) [31] equipped with a highly-sensitive and linear electron detector [32]. The microscope is combined with a Ti:Sapphire oscillator (Femtolasers Femtosource Compact) that provides us with < 15 fs laser pulses with a central wavelength of 800 nm ($\hbar\omega = 1.55$ eV) at a repetition rate of 80 MHz. We worked in a normal-incidence geometry [33] and used a Pancharatnam's phase stabilized Mach-Zehnder-interferometer [34, 35] to create pairs of mutually delayed pump and probe laser pulses with sub-femtosecond accuracy. The setup is similar to the one used in Ref. [36]. Half-wave plates in each of the two arms of the interferometer in combination with a Brewster polarization plate at the output of the interferometer were used to independently tune the power of the pump and the probe laser pulses while maintaining a common linear polarization axis. Before the laser pulses entered the microscope, the final linear polarization was adjusted to be perpendicular to a grating coupler on the sample (cf. Fig. 1A) with another, freely adjustable, half-wave plate.

The grating coupler was cut into a single-crystalline Au platelet [37] by focused ion beam milling (FIB) using a FEI Helios NanoLab 600. The sample was transferred through air into the microscope and subsequently cleaned by *in-situ* oxygen plasma etching, Argon ion sputtering, and degassing at elevated temperature in ultra-high vacuum. Prior to the photoemission experiments, we lowered the work-function of the sample by deposition of a sub-monolayer of Cs from a commercial dispenser (SAES Getters) to enable a second-order electron emission process.

**Results**

An overview of the experiment is shown in Fig. 1A. A first femtosecond laser pulse (pump pulse) excites an SPP pulse at the grating coupler. After the SPP has freely propagated for about 80 fs, a subsequent (probe) pulse forms an interaction region where the combined SPP and photon fields initiate second-order absorption liberating an electron from the surface. A direct measure of the electron emission yield provides no information about the absorbed quanta, which one might naively attribute to photon-photon, SPP-SPP, photon-SPP, or SPP-photon absorption. However, the different propagation directions of the SPP pulse and the laser pulse at the metal surface result in a significant momentum difference between them (Fig. 1B). Thus, despite being degenerate in energy, during absorption different combinations of SPPs and photons are associated with different in-plane field momenta [29, 38, 39] (Fig. 1C). As the key concept of this work, we will show that these different field momenta provide an electron emission signature that enables us to experimentally identify and separate quantum path interferences that arise from the interference of the different absorption processes that are shown in Fig. 1C.

The interaction region formed by the SPP and the probe laser pulse appears in the PEEM image as a spatial fringe modulation (Fig. 2A), which is a signature of the propagating SPP pulse at this particular pump-probe delay. In the classical field picture, this characteristic electron emission pattern [33, 40, 41] is due to the interference of the SPP and the probe laser field. In a quantum description, however, such a fringe modulation must be attributed to quantum path interferences in the electron emission process, as shown in a recent experiment [42] on spin-orbit mixing of SPPs with orbital angular momentum [43] and circularly polarized light. Since the period length of the fringe pattern is determined by the SPP wavelength $\lambda_\text{SPP}$ the pattern is commonly referred to as a "direct conceptual visualization" of the SPP pulse [33, 44]. A profile taken through the fringe pattern, however, shows a distinct non-linearity (Fig. 2B), that appears as a second-order cross-correlation of the pulses. In Fig. 2C, we decompose this profile into contributions arising from the featureless envelopes of the pulses, contributions with periods equal to the SPP wavelength, and one contribution with half the SPP wavelength. These contributions to the nonlinear profile are a direct manifestation of different quantum path interferences that occur in the electron emission.

To understand the origin of the quantum path interferences, we first consider the probability amplitude $S_{fi}$ for the absorption processes depicted in Fig. 1C. As we will show, these processes are combined to form the discussed quantum path interferences during electron emission. We describe the emission of an electron by the absorption of two quanta from a joint initial state $|\psi\rangle$ of the probe laser and the SPP, arriving in a joint final state $|\phi\rangle$ of the fields. Based on the theory of two-photon absorption by Mollow [45] the probability amplitude can be written as

$$S_{fi}(\mathbf{R}) = \int\int dt_1 dt_2\, \mathcal{L}_{fi}(t_1, t_2) \langle\phi| \left(\hat{\mathbf{E}}^{(+)}_\text{Probe}(t_2) + \hat{\mathbf{E}}^{(+)}_\text{SPP}(\mathbf{R}, t_2)\right)\left(\hat{\mathbf{E}}^{(+)}_\text{Probe}(t_1) + \hat{\mathbf{E}}^{(+)}_\text{SPP}(\mathbf{R}, t_1)\right)|\psi\rangle. \tag{1}$$

The details of the second-order transition of the liberated electron and in particular the associated dipole moments are contained in the two-time dipole correlation function $\mathcal{L}_{fi}(t_1, t_2)$ [45] and are not relevant for the following discussion. Instead, we focus on the quantum transition of the fields, which is due to the annihilation of photons and SPPs by the positive-frequency field operators $\hat{\mathbf{E}}^{(+)}_\text{Probe}(t)$ and $\hat{\mathbf{E}}^{(+)}_\text{SPP}(\mathbf{R}, t)$. Only the SPP operator depends on the spatial coordinate $\mathbf{R}$ due to the normal incidence of the laser pulses [33]. On expanding the terms in Eq. 1 we can find all time-ordered combinations of consecutive absorptions from both fields, such as $\hat{\mathbf{E}}_{\text{Probe}}^{(+)}\hat{\mathbf{E}}_{\text{Probe}}^{(+)}$, $\hat{\mathbf{E}}_{\text{SPP}}^{(+)}\hat{\mathbf{E}}_{\text{Probe}}^{(+)}$, $\hat{\mathbf{E}}_{\text{Probe}}^{(+)}\hat{\mathbf{E}}_{\text{SPP}}^{(+)}$, and $\hat{\mathbf{E}}_{\text{SPP}}^{(+)}\hat{\mathbf{E}}_{\text{SPP}}^{(+)}$, that have been indicated in Fig. 1C. The overall probability $P$ for a second-order emission process is then given by the modulus square of the probability amplitude $S_{fi}$ summed over all final states of the fields. We arrive at

$$P(\mathbf{R}) = \int\int\int\int dt'_1 dt'_2 dt_1 dt_2 \, \mathcal{M}(t'_1, t'_2; t_1, t_2) \, \mathcal{E}(\mathbf{R}; t'_1, t'_2; t_1, t_2), \tag{2}$$

where the dipole correlation functions are absorbed into the photoelectron response function $\mathcal{M}$ as discussed in the Supplementary Note 1. The probability $P$ essentially depends on the second-order coherence function $\mathcal{E}$ of the field operators with

$$\begin{aligned}\mathcal{E}(\mathbf{R}; t'_1, t'_2; t_1, t_2) &= \left\langle \left(\hat{\mathbf{E}}_{\text{Probe}}^{(-)}(t'_1) + \hat{\mathbf{E}}_{\text{SPP}}^{(-)}(\mathbf{R}, t'_1)\right)\left(\hat{\mathbf{E}}_{\text{Probe}}^{(-)}(t'_2) + \hat{\mathbf{E}}_{\text{SPP}}^{(-)}(\mathbf{R}, t'_2)\right)\right. \\ &\left.\times \left(\hat{\mathbf{E}}_{\text{Probe}}^{(+)}(t_2) + \hat{\mathbf{E}}_{\text{SPP}}^{(+)}(\mathbf{R}, t_2)\right)\left(\hat{\mathbf{E}}_{\text{Probe}}^{(+)}(t_1) + \hat{\mathbf{E}}_{\text{SPP}}^{(+)}(\mathbf{R}, t_1)\right)\right\rangle_{\hat{\sigma}}.\end{aligned} \tag{3}$$

Here, $\langle ... \rangle_{\hat{\sigma}} = \text{Tr}(\hat{\sigma} ...)$ denotes an expectation value with respect to the joint initial state density matrix $\hat{\sigma}$ of the probe field and the SPP field, and $\hat{\mathbf{E}}_i^{(-)} = \left(\hat{\mathbf{E}}_i^{(+)}\right)^\dagger$ are the negative-frequency field (i.e., creation) operators. The second-order emission probability depends on products of annihilation and creation operators, as in cross-terms like $\hat{\mathbf{E}}_{\text{Probe}}^{(-)}\hat{\mathbf{E}}_{\text{Probe}}^{(-)}\hat{\mathbf{E}}_{\text{Probe}}^{(+)}\hat{\mathbf{E}}_{\text{SPP}}^{(+)}$, which represent quantum path interferences that are formed by the different absorption processes in Fig. 1C.

The spatial fringe modulation with the SPP wavevector $\mathbf{k}_{\text{SPP}}$ observed in the electron yield is obtained by considering the (approximate) plane-wave nature of the SPP in the surface plane $\hat{\mathbf{E}}_{\text{SPP}}^{(\pm)} \propto e^{\pm i \mathbf{k}_{\text{SPP}} \cdot \mathbf{R}}$. The terms involving products of dissimilar operators, $\hat{\mathbf{E}}_{\text{SPP}}^{(\pm)}\hat{\mathbf{E}}_{\text{SPP}}^{(\mp)}$, are independent of the SPP wavevector, those involving just one SPP field operator $\hat{\mathbf{E}}_{\text{SPP}}^{(\pm)}$ depend on $e^{\pm i \mathbf{k}_{\text{SPP}} \cdot \mathbf{R}}$, and the terms involving only products of similar operators, $\hat{\mathbf{E}}_{\text{SPP}}^{(\pm)}\hat{\mathbf{E}}_{\text{SPP}}^{(\pm)}$, depend on $e^{\pm i 2\mathbf{k}_{\text{SPP}} \cdot \mathbf{R}}$. On this basis, the measured electron emission yield profile of Fig. 2B can be decomposed into the different contributions of Fig. 2C. They arise from different mixings of the fields during electron emission, which in turn depend on integer multiples of the SPP wavevector.

To clarify which of the contributions to the electron yield must be interpreted as quantum path interferences, we expand Eq. 2 to obtain the quantum-mechanical electron emission rate in a momentum space that is spanned by the real-space periodic modulations of the electron yield. This momentum space must not be confused with the momentum space spanned by the emission angles of the liberated electrons. The resulting electron emission rate in momentum space is (see Supplementary Note 1 for a detailed derivation and discussion)

$$\Gamma_{2PPE}^{quantum}(\mathbf{K}, \Delta t) \propto \delta(\mathbf{K}) \left( \frac{\langle \hat{a}^\dagger \hat{a}^\dagger \hat{a} \hat{a} \rangle_{\hat{\sigma}}}{\mathcal{N}_{Probe}^2} + \epsilon_{SPP}^4 \frac{\langle \hat{b}^\dagger \hat{b}^\dagger \hat{b} \hat{b} \rangle_{\hat{\sigma}}}{\mathcal{N}_{SPP}^2} \right)$$

$$+ \delta(\mathbf{K}) \frac{1}{\mathcal{N}_{Probe} \mathcal{N}_{SPP}} \left( \langle \hat{a}^\dagger \hat{b}^\dagger \hat{a} \hat{b} \rangle_{\hat{\sigma}} + \langle \hat{b}^\dagger \hat{a}^\dagger \hat{b} \hat{a} \rangle_{\hat{\sigma}} + \epsilon_{SPP}^2 (\langle \hat{b}^\dagger \hat{a}^\dagger \hat{a} \hat{b} \rangle_{\hat{\sigma}} + \langle \hat{a}^\dagger \hat{b}^\dagger \hat{b} \hat{a} \rangle_{\hat{\sigma}}) \right)$$

$$+ \delta(\mathbf{K} - \mathbf{k}_{SPP}) \frac{e^{-i\omega \Delta t}}{\sqrt{\mathcal{N}_{Probe} \mathcal{N}_{SPP}}} \left( \frac{\langle \hat{a}^\dagger \hat{a}^\dagger \hat{a} \hat{b} \rangle_{\hat{\sigma}} + \langle \hat{a}^\dagger \hat{a}^\dagger \hat{b} \hat{a} \rangle_{\hat{\sigma}}}{\mathcal{N}_{Probe}} \right.$$
$$\left. + \epsilon_{SPP}^2 \frac{\langle \hat{a}^\dagger \hat{b}^\dagger \hat{b} \hat{b} \rangle_{\hat{\sigma}} + \langle \hat{b}^\dagger \hat{a}^\dagger \hat{b} \hat{b} \rangle_{\hat{\sigma}}}{\mathcal{N}_{SPP}} \right) \quad (4)$$

$$+ \delta(\mathbf{K} + \mathbf{k}_{SPP}) \frac{e^{i\omega \Delta t}}{\sqrt{\mathcal{N}_{Probe} \mathcal{N}_{SPP}}} \left( \frac{\langle \hat{b}^\dagger \hat{a}^\dagger \hat{a} \hat{a} \rangle_{\hat{\sigma}} + \langle \hat{a}^\dagger \hat{b}^\dagger \hat{a} \hat{a} \rangle_{\hat{\sigma}}}{\mathcal{N}_{Probe}} \right.$$
$$\left. + \epsilon_{SPP}^2 \frac{\langle \hat{b}^\dagger \hat{b}^\dagger \hat{b} \hat{a} \rangle_{\hat{\sigma}} + \langle \hat{b}^\dagger \hat{b}^\dagger \hat{a} \hat{b} \rangle_{\hat{\sigma}}}{\mathcal{N}_{SPP}} \right)$$

$$+ \delta(\mathbf{K} - 2\mathbf{k}_{SPP}) e^{-2i\omega \Delta t} \frac{\langle \hat{a}^\dagger \hat{a}^\dagger \hat{b} \hat{b} \rangle_{\hat{\sigma}}}{\mathcal{N}_{Probe} \mathcal{N}_{SPP}} + \delta(\mathbf{K} + 2\mathbf{k}_{SPP}) e^{2i\omega \Delta t} \frac{\langle \hat{b}^\dagger \hat{b}^\dagger \hat{a} \hat{a} \rangle_{\hat{\sigma}}}{\mathcal{N}_{Probe} \mathcal{N}_{SPP}}.$$

Here, $\mathbf{K}$ is the wavevector in the surface plane and $\epsilon_{SPP}^2$ is the squared magnitude of the SPP polarization vector. The field operators have been decomposed into annihilation and creation operators for the probe photons, $\hat{a}$ and $\hat{a}^\dagger$, and likewise for the SPPs, $\hat{b}$ and $\hat{b}^\dagger$, where $\mathcal{N}_{Probe}$ and $\mathcal{N}_{SPP}$ are the associated normalization constants.

While Eq. 4 consists of 16 different fourth-order initial state correlation functions of the SPP and the probe field, only 6 of these correlation functions are independent. The correlation functions for positive and negative wavevectors, i.e., the correlation functions in the third and fourth line of Eq. 4 as well as the two correlation functions in the last line of Eq. 4, are identical up to complex conjugation. Moreover, some of the remaining correlation functions are identical up to equal-time commutations of the involved operators. Each of these correlation functions is interpreted as an individual electron emission pathway in Liouville space [46] and we will identify which of these pathways correspond to quantum path interferences in the context of Fig. 4.

First, however, we compare our experiment to Eq. 4 and demonstrate the existence of all of the 6 independent Liouville pathways in the experimental data. For this purpose, we calculate the wavevector spectrum of the data (Fig. 2D-E) via a spatial Fourier transformation of the PEEM image of Fig. 2A. The wavevector spectrum consists of 5 distinct peaks at multiples of the SPP wavenumber $|\mathbf{k}_{SPP}| = 2\pi/\lambda_{SPP}$, located on a line perpendicular to the SPP phase fronts. The first-order peaks at $\mathbf{K} = \pm \mathbf{k}_{SPP}$ correspond to the fringe modulation of the electron yield in real space with periodicity $\lambda_{SPP}$, justifying the usual interpretation of the fringe pattern as a "direct conceptual visualization" of the SPP pulse. The second-order peaks at $\mathbf{K} = \pm 2\mathbf{k}_{SPP}$, however, correspond to a periodic modulation of the electron yield at half the SPP wavelength and are a direct consequence of the nonlinear emission process. We do not observe third order peaks at $\mathbf{K} = \pm 3\mathbf{k}_{SPP}$, which corroborates that the electron emission process is of second order.

Each of the different Liouville pathways in Eq. 4 is characterized by a distinct wavevector $\mathbf{K}$, a distinct harmonic delay-dependence, and a distinct dependence on the product of annihilation and creation operators. The quantum-mechanical transition rate closely resembles the structure of the common phenomenological model [40, 47], which can be obtained from Eqs. 2-4 using the correspondence principle. In the resulting classical field approximation, each of the annihilation and creation operators contributes to the correlation functions with the square-root of the intensities of the respective fields, i.e., the field-strengths.

This relationship allows us to sort the contributions of the 6 independent Liouville pathways by the powers of their field-strengths and their signatures in the electron yield in momentum space (Fig. 2E). The second-order peaks at $\mathbf{K} = \pm 2\mathbf{k}_{\text{SPP}}$ exclusively consist of a contribution proportional to $E_{\text{SPP}}^2 E_{\text{Probe}}^2$. The first-order peaks at $\mathbf{K} = \pm \mathbf{k}_{\text{SPP}}$, however, originate from two contributions: one is proportional to $E_{\text{SPP}} E_{\text{Probe}}^3$ and the other one is proportional to $E_{\text{SPP}}^3 E_{\text{Probe}}$. The situation is even more complicated for the central peak at $\mathbf{K} = 0$: it consists of three contributions, proportional to $E_{\text{SPP}}^4$, $E_{\text{Probe}}^4$, and $E_{\text{SPP}}^2 E_{\text{Probe}}^2$. Note that the contributions proportional to $E_{\text{SPP}}^4$ and to $E_{\text{Probe}}^4$ are not shown in Figs. 2c and 2f as these only correspond to spatially broad plasmoemission- [21] and photoemission backgrounds, respectively.

The different scaling behavior of the Liouville pathways with the field amplitudes provides a means to disentangle them experimentally. We systematically change the pump and probe powers to independently control the absorption probability from the SPP field and the probe laser field, respectively. This procedure provides information on which of the two fields and pathways dominate the electron emission. Figure 3 shows the measured integral amplitude of each wavevector peak as a function of the normalized probe field strength, i.e., the normalized square-root of the probe power. We repeated the measurement for 4 representative pump powers to vary the field strength of the SPP. Each of the curves was normalized to its maximal integral amplitude. The results are plotted on a double-logarithmic scale in Fig. 3 such that power laws appear as straight lines with respective slopes. In Fig. 3A the integral amplitude of the second-order peaks at $\mathbf{K} = \pm 2\mathbf{k}_{\text{SPP}}$ depends for all pump powers on the normalized probe field-strength as $E_{\text{SPP}}^2 E_{\text{Probe}}^2$. Figure 3b shows the integral amplitude of the first-order peaks at $\mathbf{K} = \pm \mathbf{k}_{\text{SPP}}$. We find that for low pump powers, i.e., weak SPP excitation, the amplitude is dominated by the contribution proportional to $E_{\text{SPP}} E_{\text{Probe}}^3$ with a slope of three. This characteristic motivates the recently reported vector microscopy [36]. For high pump powers, i.e., strong SPP excitation, the amplitude in Fig. 3B becomes dominated by the contribution proportional to $E_{\text{SPP}}^3 E_{\text{Probe}}$ with a slope of one. Figure 3c shows the results for the central wavevector peak at $\mathbf{K} = 0$. As this central peak also contains all long-range background modulations of the TR-PEEM images, such as plasmo- and photoemission backgrounds, we subtracted a probe-power-independent constant from each of the curves in Fig. 3C (see Supplementary Note 2 for details). Note that this subtracted constant includes the contribution proportional to $E_{\text{SPP}}^4$. For low pump powers, i.e., weak SPP excitation, the amplitude of the central peak is dominated by the contribution proportional to $E_{\text{Probe}}^4$. As we increase the pump power, the amplitude of the central peak becomes dominated by the contribution proportional to $E_{\text{SPP}}^2 E_{\text{Probe}}^2$.

In a classical particle picture, one would expect the probability for an electron to absorb a photon or an SPP to depend on the intensities of the respective fields, which are proportional to the squared magnitudes of the field strengths. This expectation implies it must be possible to attribute a number of absorbed quanta from each of the involved fields to every liberated electron. However, in this classical particle picture it is difficult to interpret the experimental existence of contributions to the electron yield like $E_{\text{SPP}} E_{\text{Probe}}^3$ that scale as odd powers of the field-strengths. Such difficulties are not encountered in the purely quantum analysis leading to Eq. (4).

After having demonstrated the existence of all Liouville pathways in Eq. 4 in the experimental data, we now identify which of these pathways must be interpreted as quantum path interferences that arise from the interference of the SPP- and photon absorption processes in Fig. 1C. All Liouville pathways for the electron emission process in Eq. 4 (except for complex conjugate pathways) are summarized in analogy to double-sided Feynman diagrams [30] in Fig. 4. Each of the pathways in Fig. 4 consist of four arrows, where the colors red and blue represent photon absorption and SPP absorption, respectively. The arrows on the left side of each pathway correspond to creation operators, the ones on the right side correspond to annihilation operators. Each of the depicted Liouville pathways is associated with a momentum equal to the momentum difference between the left- and right-hand side, which gives rise to the respective peaks in the wavevector spectrum in Fig. 2D.

The central wavevector peak at $\mathbf{K} = 0$ arises from electron emission by the consecutive absorption of two probe photons (pathway (A)), the consecutive absorption of two SPP quanta (pathway (B)), as well as cooperative pathways (C) – (E). While pathway (C) and (D) correspond to the consecutive absorption of each an SPP quantum and a probe photon, pathway (E) corresponds to the interference of the consecutive absorption of an SPP quantum and a probe photon with the respective inversely-ordered process (non-oscillatory double-mixing [42]). It is worth noting, however, that within the approximations in Supplementary Note 1 the pathways (C), (D) and (E) are physically equivalent, as they can be transformed into each other via trivial commutations of creation or annihilation operators. Thus, all Liouville pathways that contribute to the central wavevector peak only depend on the SPP and probe photon populations.

The remaining Liouville pathways (F) – (J) cannot be explained by the simple consecutive absorption of SPPs and probe photons but instead must be interpreted as quantum path interferences of fundamentally different electron emission pathways. In the probe-dominated pathways (F) and (G), the consecutive absorption of two probe photons interferes with the consecutive absorption of an SPP and a probe photon. This situation is reversed for the SPP-dominated pathways (H) and (I), where instead the consecutive absorption of two SPPs interferes with the consecutive absorption of an SPP and a probe photon. Moreover, the second-order wavevector peak at $\mathbf{K} = \pm 2\mathbf{k}_{SPP}$ consists exclusively of the interference of the consecutive absorption of two SPPs with the consecutive absorption of two probe photons (pathway (J)). The single-mixing pathways of the first-order wavevector peak and the double-mixing pathway of the second-order wavevector peak probe the mutual first and second-order coherences of the SPPs and of the probe photons, respectively.

It is remarkable that some of the discussed Liouville pathways result in observable quantum path interferences in the electron emission - a consequence of the nonlinear mixing of the fields in the emission process. It is important to note that by utilizing momentum resolution we could resolve which quantum path interferences (Fig. 4F-J) contribute to the electron emission, but we did not resolve the individual absorption processes (as in Fig. 1C) that constitute the quantum path interferences. Resolving the individual absorption processes would be the goal of a which-way experiment, and doing so would indeed destroy the observed quantum path interferences.

**Discussion**

Quantum path interferences are a manifestation of the inherent quantum nature of fundamental interactions. Our approach to electron emission in the simultaneous presence of SPPs and light confirms that Liouville pathways can be disentangled by their power-dependent contributions in a momentum space that consists of discrete spots. Addressing more complex, non-trivial quantum correlations between light and SPPs, like in entangled SPP-photon pairs [48], constitutes the natural progression of our work. We believe that interferences between transitions involving additional quantum numbers for the SPPs, such as spin- and orbital angular momentum [42, 43, 49], can be studied most effectively in momentum space as well. Ultimately, adding energy resolution and electron momentum resolution to our technique will provide a route to study non-trivial quantum correlations between interacting quantum electrons, quantum light, and quantum SPPs in the future.

**Acknowledgments:** We thank Frank Jahnke and Christopher Gies for discussion about the theory. We further thank Bettina Frank for providing us with high-quality sample substrates.

**Funding:** This work was funded by the Deutsche Forschungsgemeinschaft (DFG, German Research Foundation) through Collaborative Research Center SFB1242 "Non-equilibrium dynamics of condensed matter in the time domain" (Project-ID 278162697), as well as SPP1839 and GRK2642. The Stuttgart group is supported by ERC AdG ComplexPlas and PoC 3DPrintedoptics.

**Author contributions:** P.D., D.J., and F.-J.M.z.H. did the time-resolved PEEM experiments. P.D. performed the data analysis and developed the theory. All authors contributed to the data interpretation and discussions. The manuscript was written through contributions of all authors.

**Competing interests:** All authors declare they have no competing interests.

**Data and materials availability:** All data needed to evaluate the conclusions in the paper are present in the paper and/or the Supplementary Materials. Additional data related to this paper may be requested from the authors.

**Figures and Tables**

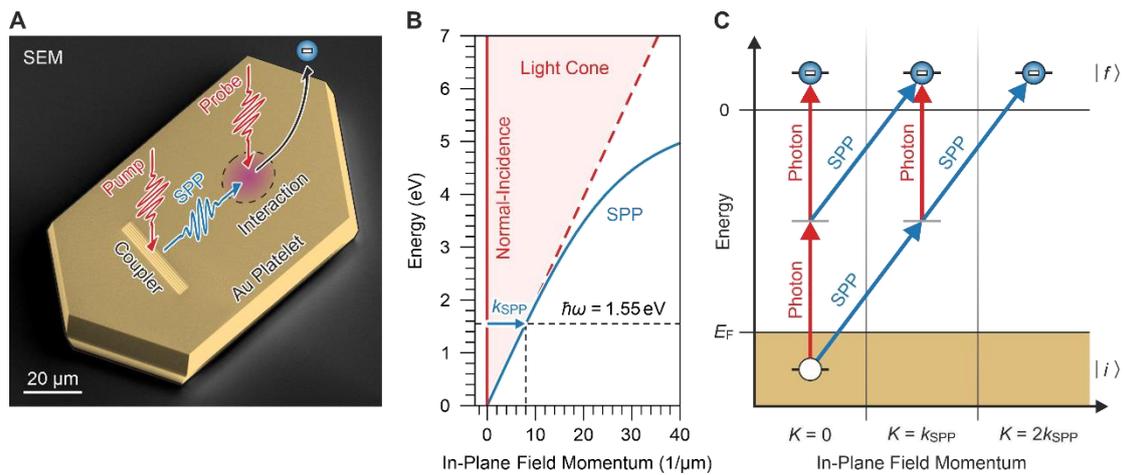

**Fig. 1. Mixing of SPPs and light in electron emission.**
(**A**) Sketch of the utilized pump-probe scheme. The scanning electron micrograph shows a platelet similar to the one used for the presented experiments. The arrows illustrate the different light and SPP pulses. (**B**) Dispersion relation for SPPs and light as a function of the momentum in the surface plane. It is the momentum-mismatch between normally-incident light and the SPPs that we exploit to distinguish individual quantum path interferences. (**C**) Energy-level diagram of the different second-order electron emission pathways that can occur in the interaction region of SPPs and light. The different states and paths are sorted by the in-plane momentum transfer during the emission process.

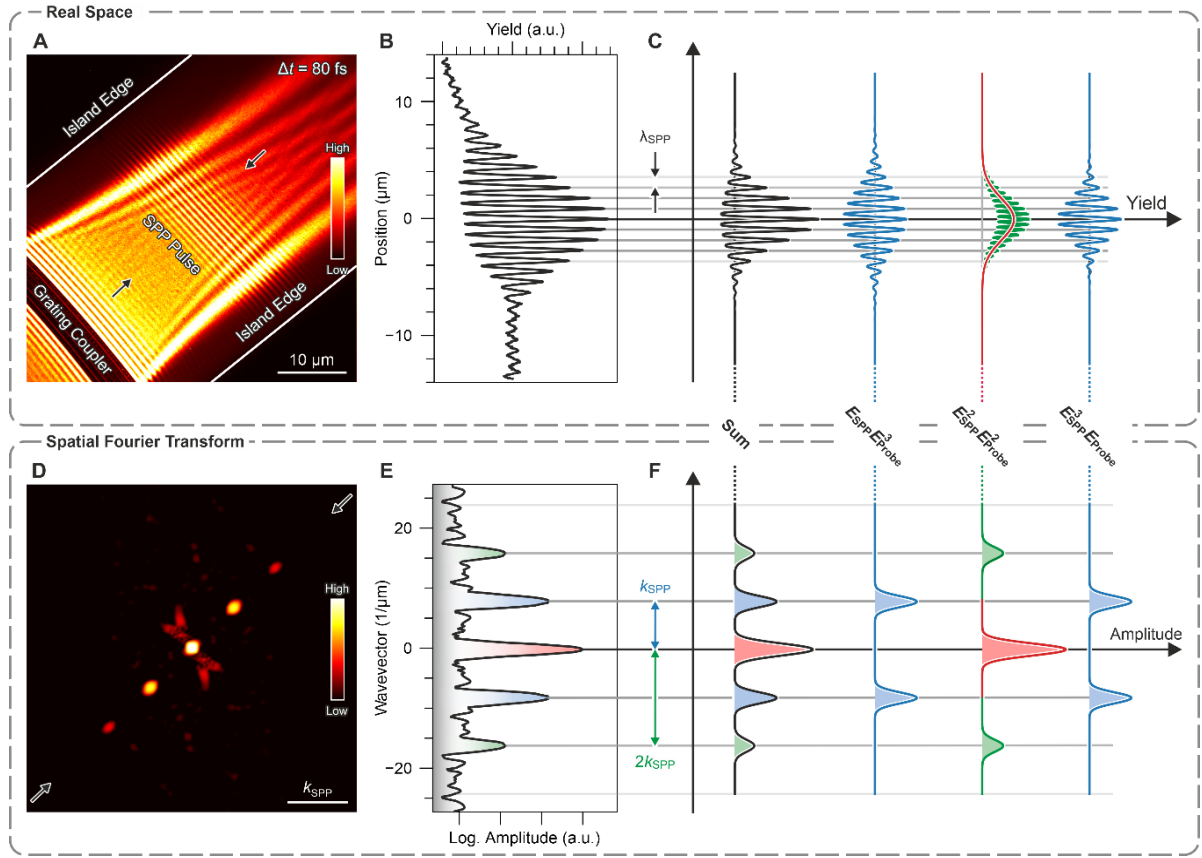

**Fig. 2. Fourier decomposition of SPP-light correlations in PEEM.**
**(A)** Time-resolved PEEM image of an SPP pulse 80 fs after excitation at the grating coupler, depicted in a linear false-color scale. The fringes in the center of the image are a direct conceptual visualization of the propagating SPP pulse. **(B)** Section through the fringe-pattern in the electron yield distribution in real-space as indicated by the arrows in **(A)**. **(C)** Sketch of the contribution of the different terms in Eq. 4 to the spatial fringe-pattern in the electron yield **(B)** sorted by powers in the SPP- and the probe field. **(D)** Wavevector spectrum computed via a windowed Fourier transform of an electron-optically magnified image of the interaction region in **(A)** depicted on a logarithmic false-color scale. The five visible peaks arise from the fringe modulation in real-space, and their equidistant spacing is given by the SPP wavenumber. **(E)** Section through the SPP wavevector spectrum as indicated by the arrows in **(D)**. **(F)** Sketch of the contribution of the different terms in Eq. 4 to the wavevector spectrum in **(E)** sorted by powers in the SPP- and the probe field.

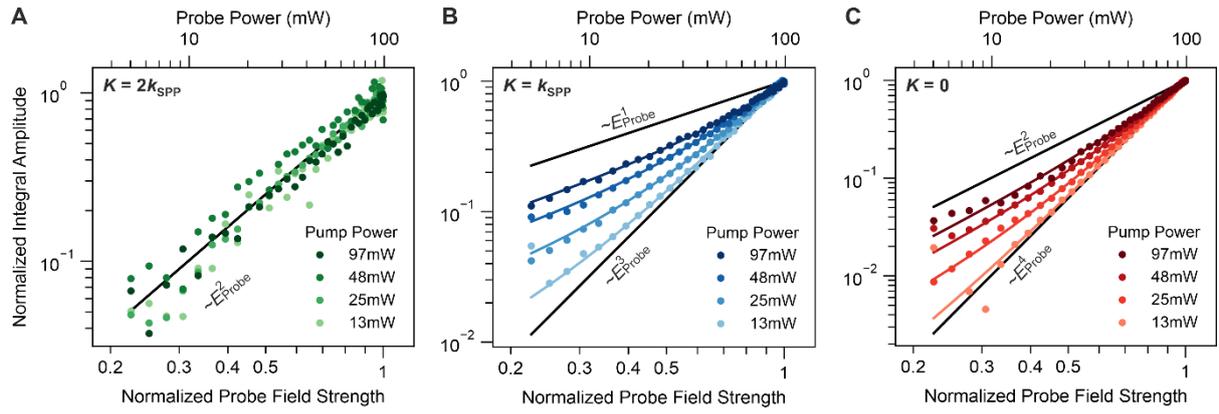

**Fig. 3. Separation of wavevector peak contributions.**
(**A**) Integral amplitude of the second-order wavevector peak at $K = \pm 2k_{SPP}$, (**B**) the first-order wavevector peak at $K = \pm k_{SPP}$, (**C**) the central wavevector peak at $K = 0$ as a function of the normalized probe field strength for four different pump powers, plotted on a double logarithmic scale. The measurements match the power laws expected from Eq. 2 well, as depicted by the straight black lines.

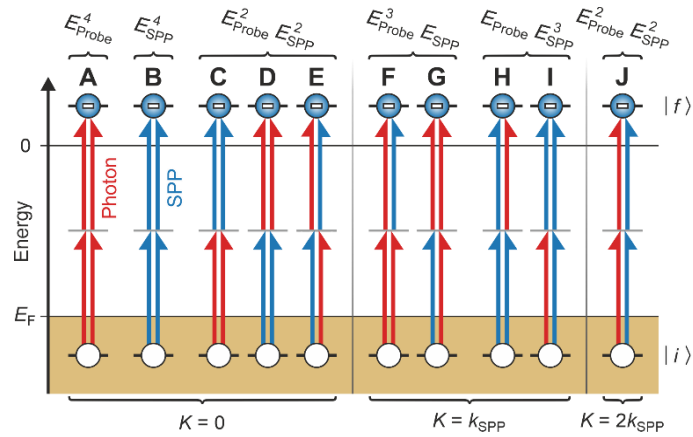

**Fig. 4. Microscopic picture of the quantum pathways during electron emission.**
An electron is liberated from an initial state |i⟩ below the Fermi energy $E_F$ to a final state |f⟩ in the vacuum by the second-order absorption of photons (red arrows) and SPPs (blue arrows). In each of the diagrams, the pathway given by the two arrows on the left-hand side interferes with the pathway given by the arrows on the right-hand side of the respective diagram. The diagrams are grouped by their dependence on the strength of the SPP and the probe field and by their wavevector contribution.

# Momentum space separation of quantum path interferences between photons and surface plasmon polaritons in nonlinear photoemission microscopy

Supplementary Information

Pascal Dreher[1], David Janoschka[1], Harald Giessen[2], Ralf Schützhold[1,4,5], Timothy J. Davis[1,2,3], Michael Horn-von Hoegen[1], and Frank-J. Meyer zu Heringdorf[1,6]

[1]Faculty of Physics and Center for Nanointegration, Duisburg-Essen (CENIDE), University of Duisburg-Essen, 47048 Duisburg, Germany
[2]4th Physics Institute, Research Center SCoPE, and Integrated Quantum Science and Technology Center, University of Stuttgart, 70569 Stuttgart, Germany
[3]School of Physics, University of Melbourne, Parkville, Victoria 3010, Australia
[4]Helmholtz-Zentrum Dresden-Rossendorf, Bautzner Landstrasse 400, 01328 Dresden, Germany
[5]Institut für Theoretische Physik, Technische Universität Dresden, 01062 Dresden, Germany
[6]Interdisciplinary Center for the Analytics on the Nanoscale (ICAN), 47057 Duisburg, Germany

# Supplementary Note 1: Calculation of the electron yield

The image-contrast mechanism for the observation of surface plasmon polaritons (SPPs) in interferometric time-resolved two-photon photoemission electron microscopy is typically described in a phenomenological semi-classical model [1]. This is, however, not sufficient to interpret the results from the main text in a particle picture, as the semi-classical model only considers the electric fields and thus fundamentally lacks the concept of absorption of individual photons or SPPs.

In the following we will first briefly review the semi-classical field-based description and generalize it to our Fourier-space approach to photoemission electron microscopy from the main text. We will then based on the theory of two-photon absorption derive a quantum-mechanical generalization of the established semi-classical model. The derived model will as well be discussed in Fourier-space and its connection to the experimentally varied parameters will be highlighted.

## 1.1 Electric fields

To model the observed electron yield, either by a semiclassical approach or by a quantum-mechanical approach, we first need to know the functional form of all relevant fields at the surface. The investigated sample is modeled as a single vacuum-gold interface located at $z = 0$. The dielectric function of the problem is given by $\varepsilon(z > 0) = 1$ in vacuum and by $\varepsilon(z \leq 0) = \varepsilon_m$ in the gold platelet.

As discussed in the main text we approximate the electric field of the probe laser pulse $\mathbf{E}_{\text{Probe}}(\mathbf{r}, t)$, which is normally-incident from vacuum onto the surface, as a monochromatic wave. Naively, one could write for the electric field of the probe laser

$$\mathbf{E}_{\text{Probe}}(\mathbf{r}, t) = E_{\text{Probe}} \boldsymbol{\epsilon}_{\text{Probe}} e^{-ikz} e^{-i\omega t}. \tag{S1}$$

Here, $E_{\text{Probe}}$ is the amplitude, $\omega$ is the frequency, and $k = \omega/c_0$ is the wavenumber of the probe field with $c_0$ being the speed of light in vacuum. The polarization vector of the probe laser pulse $\boldsymbol{\epsilon}_{\text{Probe}}$ lies in the $(x,y)$-plane and is assumed to be normalized, i.e. $|\boldsymbol{\epsilon}_{\text{Probe}}| = 1$.

It is clear that Eq. (S1) only holds in free space. More generally, the probe field is partially reflected and partially transmitted at the surface. The transmitted part of the probe field furthermore is damped and screened in the gold platelet. Electrons are, however, exclusively liberated by absorption of quanta from the local electric field within the gold platelet. While the reflected part of the probe field can also interact with liberated electrons [2], such effects are not of relevance here. Usually, the Fresnel equations are employed in photoemission to model the effect of the surface onto the electric field that liberates the electrons [3]–[5] and it has been shown that this treatment is valid down to atomic length scales [6]. To model the partial transmission and subsequent damping of the probe field in the gold platelet we introduce a mode function

$$u_{\text{Probe}}(\mathbf{r}) = \begin{cases} e^{-ikz} + \mathcal{R}e^{ikz} & z > 0 \\ \mathcal{T}e^{z/\delta_e} e^{-ik'z} & z \leq 0 \end{cases}, \tag{S2}$$

where $\mathcal{R}$ and $\mathcal{T}$ are the complex reflection and transmission amplitudes of the probe field given by the Fresnel equations, $k'$ is the wavenumber and $\delta_e$ is the penetration depth of the probe field in the gold platelet. Since the probe field is normally incident onto the surface and birefringence is negligible for gold, the probe polarization does not change upon transmission at the surface, which can easily be shown using the Fresnel equations. This allows us to write the probe field as

$$\mathbf{E}_{\text{Probe}}(\mathbf{r}, t) = E_{\text{Probe}} \boldsymbol{\epsilon}_{\text{Probe}} u_{\text{Probe}}(\mathbf{r}) e^{-i\omega t}. \tag{S3}$$

Just like the probe field, we approximate the SPP pulse by a single plane SPP wave. The electric field of a plane SPP wave for an isotropic and non-magnetic interface, like for our vacuum-gold interface, can be written as

$$\mathbf{E}_{\text{SPP}}(\mathbf{r}, t) = E_{\text{SPP}} \boldsymbol{\epsilon}_{\text{SPP}}(z) u_{\text{SPP}}(\mathbf{r}) e^{-i\omega t}, \tag{S4}$$

where $\mathbf{r}$ is the position, $t$ is the time, $E_{\text{SPP}}$ is the amplitude, $\boldsymbol{\epsilon}_{\text{SPP}}$ is the polarization vector, and $u_{\text{SPP}}(\mathbf{r})$ is the mode function. In our experiment, the frequency of the SPP $\omega$ is identical to the frequency of the probe, since the SPP is resonantly excited by the same laser delivering the probe pulses. The mode function of the SPP field has the simple form

$$u_{\text{SPP}}(\mathbf{r}) = e^{i\mathbf{k}_{\text{SPP}} \cdot \mathbf{r}} e^{-\gamma_{\text{SPP}}(z)|z|}, \tag{S5}$$

which corresponds to a plane wave in the ($x,y$)-plane of the interface, that exponentially decays along the $z$-direction. Here, $\mathbf{k}_{\text{SPP}}$ is the in-plane wavevector of the SPP and $\gamma_{\text{SPP}}(z)$ is the decay constant (in $z$-direction) of the SPP field. The SPP wavenumber is given by

$$|\mathbf{k}_{\text{SPP}}| = k_{\text{SPP}} = k \sqrt{\frac{\varepsilon_m}{1 + \varepsilon_m}} \tag{S6}$$

and the decay constant of the $z$-component of the SPP's electric field $\gamma_{\text{SPP}}(z)$ is related to the SPP wavenumber by $\gamma_{\text{SPP}}(z)^2 = k_{\text{SPP}}^2 - \varepsilon(z)k^2$. The polarization vector of the SPP is given by

$$\boldsymbol{\epsilon}_{\text{SPP}}(z) = \frac{\mathbf{k}_{\text{SPP}}}{k_{\text{SPP}}} \mp i \frac{k_{\text{SPP}}}{\gamma_{\text{SPP}}(z)} \mathbf{z} = \boldsymbol{\epsilon}_{\text{SPP}}^{\parallel} + \boldsymbol{\epsilon}_{\text{SPP}}^{\perp}(z), \tag{S7}$$

where $\mathbf{z}$ is the unit vector along the $z$-direction and thus perpendicular to the interface, and $\boldsymbol{\epsilon}_{\text{SPP}}^{\parallel}$ and $\boldsymbol{\epsilon}_{\text{SPP}}^{\perp}(z)$ are the in-plane and out-of-plane component of the SPP polarization vector, respectively. The negative sign of the $z$-component holds for $z > 0$, while the positive sign holds for $z \leq 0$. Note, that the $z$-component of the SPP polarization vector is discontinuous across the interface. The polarization vector is thus not normalized and the absolute value of the SPP field is not given by the amplitude of the field. Rather, the in-plane component of the polarization vector, which in contrast is continuous, is normalized and the amplitude of the field gives the amplitude of the in-plane component of the SPP field vector.

## 1.2 Semiclassical Approach

In a semiclassical simplification, we can calculate the second-order electron yield $Y_{\text{2PPE}}(\mathbf{R}, \Delta t)$ at position $\mathbf{R}$ in the surface plane and at pump-probe delay $\Delta t$ as [1]

$$Y_{\text{2PPE}}(\mathbf{R}, \Delta t) \propto \int_{-\infty}^{\infty} dt \left| \mathbf{E}(\mathbf{R}, t) \right|^4 \propto \int_{-\infty}^{\infty} dt \left| \mathbf{E}_{\text{SPP}}(\mathbf{R}, t) + \mathbf{E}_{\text{Probe}}(\mathbf{R}, t - \Delta t) \right|^4, \tag{S8}$$

where $\mathbf{E}(\mathbf{R}, t)$ is the overall electric field at the surface, and $\mathbf{E}_{\text{SPP}}(\mathbf{R}, t)$ and $\mathbf{E}_{\text{Probe}}(\mathbf{R}, t - \Delta t)$ are the electric field of the SPP pulse and the probe laser pulse, respectively. Strictly speaking, we would need to include the full $z$-dependence of all fields and of the electron emission probability and integrate out the $z$-coordinate. Since both fields and the electron emission probability decay exponentially in the metal, electrons dominantly originate from a small layer within the surface. For our analysis in the main text, however, we are only interested in modulations of the electron yield within the surface plane and thus only consider a contribution at $z = 0^-$, i.e. only consider electron emission from within the metal infinitesimally below the surface. By doing so we neglect a constant prefactor depending on the decay constants of the fields and the emission probability, which is not of importance for our following discussion. The resulting probe field is $\mathbf{E}_{\text{Probe}}(\mathbf{R}, t) \approx E_{\text{Probe}} \boldsymbol{\epsilon}_{\text{Probe}} e^{-i\omega t}$ and the resulting SPP field is $\mathbf{E}_{\text{SPP}}(\mathbf{R}, t) \approx E_{\text{SPP}} \boldsymbol{\epsilon}_{\text{SPP}} e^{i\mathbf{k}_{\text{SPP}} \cdot \mathbf{R}} e^{-i\omega t}$, where we absorbed the transmission coefficient for the probe field into the probe field amplitude $\mathcal{T} E_{\text{Probe}} \to E_{\text{Probe}}$ and where we set $\boldsymbol{\epsilon}_{\text{SPP}} = \boldsymbol{\epsilon}_{\text{SPP}}(z = 0^-)$.

### 1.2.1 Transition Rate

In our plane-wave approximation the explicit time-dependence of the integrand in Eq. (S8) vanishes. It is thus more instructive to work with the transition rate

$$\begin{aligned}
\Gamma_{\text{2PPE}}^{\text{classical}}(\mathbf{R}, \Delta t) &= \frac{\mathrm{d}Y_{\text{2PPE}}(\mathbf{R}, \Delta t)}{\mathrm{d}t} \\
&\propto \lim_{t \to \infty} \frac{\mathrm{d}}{\mathrm{d}t} \int_{-\infty}^{t} \mathrm{d}t' \left| \mathbf{E}_{\text{SPP}}(\mathbf{R}, t') + \mathbf{E}_{\text{Probe}}(\mathbf{R}, t' - \Delta t) \right|^4.
\end{aligned} \tag{S9}$$

Expanding the binomial in Eq. (S9) and inserting the definitions of the fields Eq. (S3) and Eq. (S4) leaves us with

$$\begin{aligned}
\Gamma_{\text{2PPE}}^{\text{classical}}(\mathbf{R}, \Delta t) \propto\ & E_{\text{Probe}}^4 |\boldsymbol{\epsilon}_{\text{Probe}}|^4 + E_{\text{SPP}}^4 |\boldsymbol{\epsilon}_{\text{SPP}}|^4 \\
& + 2 E_{\text{SPP}}^2 E_{\text{Probe}}^2 |\boldsymbol{\epsilon}_{\text{SPP}}|^2 |\boldsymbol{\epsilon}_{\text{Probe}}|^2 \left( 1 + \frac{|\boldsymbol{\epsilon}_{\text{Probe}}^* \cdot \boldsymbol{\epsilon}_{\text{SPP}}|^2}{|\boldsymbol{\epsilon}_{\text{SPP}}|^2 |\boldsymbol{\epsilon}_{\text{Probe}}|^2} \right) \\
& + 4 E_{\text{SPP}} E_{\text{Probe}} \left( E_{\text{SPP}}^2 |\boldsymbol{\epsilon}_{\text{SPP}}|^2 + E_{\text{Probe}}^2 |\boldsymbol{\epsilon}_{\text{Probe}}|^2 \right) \\
& \times \operatorname{Re} \left\{ \boldsymbol{\epsilon}_{\text{Probe}}^* \cdot \boldsymbol{\epsilon}_{\text{SPP}} e^{i(\mathbf{k}_{\text{SPP}} \cdot \mathbf{R} - \omega \Delta t)} \right\} \\
& + 2 E_{\text{SPP}}^2 E_{\text{Probe}}^2 \operatorname{Re} \left\{ (\boldsymbol{\epsilon}_{\text{Probe}}^* \cdot \boldsymbol{\epsilon}_{\text{SPP}})^2 e^{i(2\mathbf{k}_{\text{SPP}} \cdot \mathbf{R} - 2\omega \Delta t)} \right\}.
\end{aligned} \tag{S10}$$

In the main text we analyze the electron yield in reciprocal space. A Fourier transform of Eq. (S10) in the two-dimensional position coordinate $\mathbf{R}$ leaves us with the transition rate in reciprocal space

$$\begin{aligned}
\Gamma_{\text{2PPE}}^{\text{classical}}(\mathbf{K}, \Delta t) \propto\; & (2\pi)^2 \delta(\mathbf{K}) \left( E_{\text{Probe}}^4 |\boldsymbol{\epsilon}_{\text{Probe}}|^4 + E_{\text{SPP}}^4 |\boldsymbol{\epsilon}_{\text{SPP}}|^4 \right) \\
& + 2(2\pi)^2 \delta(\mathbf{K}) E_{\text{SPP}}^2 E_{\text{Probe}}^2 |\boldsymbol{\epsilon}_{\text{SPP}}|^2 |\boldsymbol{\epsilon}_{\text{Probe}}|^2 \left( 1 + \frac{|\boldsymbol{\epsilon}_{\text{Probe}}^* \cdot \boldsymbol{\epsilon}_{\text{SPP}}|^2}{|\boldsymbol{\epsilon}_{\text{SPP}}|^2 |\boldsymbol{\epsilon}_{\text{Probe}}|^2} \right) \\
& + 2(2\pi)^2 E_{\text{SPP}} E_{\text{Probe}} \left( E_{\text{SPP}}^2 |\boldsymbol{\epsilon}_{\text{SPP}}|^2 + E_{\text{Probe}}^2 |\boldsymbol{\epsilon}_{\text{Probe}}|^2 \right) \\
& \times \left( \boldsymbol{\epsilon}_{\text{Probe}}^* \cdot \boldsymbol{\epsilon}_{\text{SPP}} \delta(\mathbf{K} - \mathbf{k}_{\text{SPP}}) e^{-i\omega \Delta t} + \boldsymbol{\epsilon}_{\text{Probe}} \cdot \boldsymbol{\epsilon}_{\text{SPP}}^* \delta(\mathbf{K} + \mathbf{k}_{\text{SPP}}) e^{i\omega \Delta t} \right) \\
& + (2\pi)^2 E_{\text{SPP}}^2 E_{\text{Probe}}^2 \Big( (\boldsymbol{\epsilon}_{\text{Probe}}^* \cdot \boldsymbol{\epsilon}_{\text{SPP}})^2 \delta(\mathbf{K} - 2\mathbf{k}_{\text{SPP}}) e^{-2i\omega \Delta t} \\
& \hspace{3cm} + (\boldsymbol{\epsilon}_{\text{Probe}} \cdot \boldsymbol{\epsilon}_{\text{SPP}}^*)^2 \delta(\mathbf{K} + 2\mathbf{k}_{\text{SPP}}) e^{2i\omega \Delta t} \Big),
\end{aligned} \quad (S11)$$

where $\delta$ denotes the Dirac delta function. Eq. (S11) holds for any probe polarization and any orientation of the in-plane SPP polarization. In the presented experiments, however, we chose the probe polarization to be collinear with the in-plane SPP polarization. Recalling that we have $|\boldsymbol{\epsilon}_{\text{Probe}}| = 1$ and $|\boldsymbol{\epsilon}_{\text{SPP}}^{\parallel}| = 1$, we conclude $\boldsymbol{\epsilon}_{\text{Probe}}^* \cdot \boldsymbol{\epsilon}_{\text{SPP}} = 1$ and write for Eq. (S11)

$$\begin{aligned}
\Gamma_{\text{2PPE}}^{\text{classical}}(\mathbf{K}, \Delta t) \propto\; & (2\pi)^2 \delta(\mathbf{K}) \left( E_{\text{Probe}}^4 + E_{\text{SPP}}^4 \left( 1 + |\boldsymbol{\epsilon}_{\text{SPP}}^{\perp}|^2 \right)^2 \right) \\
& + 2(2\pi)^2 \delta(\mathbf{K}) E_{\text{SPP}}^2 E_{\text{Probe}}^2 \left( 2 + |\boldsymbol{\epsilon}_{\text{SPP}}^{\perp}|^2 \right) \\
& + 2(2\pi)^2 E_{\text{SPP}} E_{\text{Probe}} \left( E_{\text{SPP}}^2 \left( 1 + |\boldsymbol{\epsilon}_{\text{SPP}}^{\perp}|^2 \right) + E_{\text{Probe}}^2 \right) \\
& \times \left( \delta(\mathbf{K} - \mathbf{k}_{\text{SPP}}) e^{-i\omega \Delta t} + \delta(\mathbf{K} + \mathbf{k}_{\text{SPP}}) e^{i\omega \Delta t} \right) \\
& + (2\pi)^2 E_{\text{SPP}}^2 E_{\text{Probe}}^2 \left( \delta(\mathbf{K} - 2\mathbf{k}_{\text{SPP}}) e^{-2i\omega \Delta t} + \delta(\mathbf{K} + 2\mathbf{k}_{\text{SPP}}) e^{2i\omega \Delta t} \right).
\end{aligned} \quad (S12)$$

We can formulate the transition rate in Eq. (S12) also in terms of the intensities of the fields, i.e. use the identities

$$I_{\text{Probe}} = \frac{1}{2} c_{\text{Probe}} \varepsilon_0 |\mathbf{E}_{\text{Probe}}|^2 = \frac{1}{2} c_{\text{Probe}} \varepsilon_0 E_{\text{Probe}}^2 \quad (S13a)$$

$$I_{\text{SPP}} = \frac{1}{2} c_{\text{SPP}} \varepsilon_0 |\mathbf{E}_{\text{SPP}}|^2 = \frac{1}{2} c_{\text{SPP}} \varepsilon_0 E_{\text{SPP}}^2 \left( 1 + |\boldsymbol{\epsilon}_{\text{SPP}}^{\perp}|^2 \right). \quad (S13b)$$

After carrying out this substitution we arrive at

$$\begin{aligned}
\Gamma_{\text{2PPE}}^{\text{classical}}(\mathbf{K}, \Delta t) \propto{}& \left(\frac{4\pi}{\varepsilon_0}\right)^2 \delta(\mathbf{K}) \left(\frac{I_{\text{Probe}}^2}{c_{\text{Probe}}^2} + \frac{I_{\text{SPP}}^2}{c_{\text{SPP}}^2}\right) \\
& + 2\left(\frac{4\pi}{\varepsilon_0}\right)^2 \delta(\mathbf{K}) \frac{I_{\text{Probe}} I_{\text{SPP}}}{c_{\text{Probe}} c_{\text{SPP}}} \frac{2 + |\epsilon_{\text{SPP}}^\perp|^2}{1 + |\epsilon_{\text{SPP}}^\perp|^2} \\
& + 2\left(\frac{4\pi}{\varepsilon_0}\right)^2 \sqrt{\frac{I_{\text{Probe}} I_{\text{SPP}}}{c_{\text{Probe}} c_{\text{SPP}} (1 + |\epsilon_{\text{SPP}}^\perp|^2)}} \left(\frac{I_{\text{Probe}}}{c_{\text{Probe}}} + \frac{I_{\text{SPP}}}{c_{\text{SPP}}}\right) \\
& \times \left(\delta(\mathbf{K} - \mathbf{k}_{\text{SPP}}) e^{-i\omega\Delta t} + \delta(\mathbf{K} + \mathbf{k}_{\text{SPP}}) e^{i\omega\Delta t}\right) \\
& + 2\left(\frac{4\pi}{\varepsilon_0}\right)^2 \frac{I_{\text{Probe}} I_{\text{SPP}}}{c_{\text{Probe}} c_{\text{SPP}} (1 + |\epsilon_{\text{SPP}}^\perp|^2)} \\
& \times \left(\delta(\mathbf{K} - 2\mathbf{k}_{\text{SPP}}) e^{-2i\omega\Delta t} + \delta(\mathbf{K} + 2\mathbf{k}_{\text{SPP}}) e^{2i\omega\Delta t}\right),
\end{aligned} \qquad (S14)$$

which is our final result for the semiclassical second-order transition rate in reciprocal space.

## 1.3 Quantum-Mechanical Approach

We base our quantum-mechanical calculation of the second-order electron yield on the theory of two-photon absorption developed by Mollow [7]. In this approach, the probability $P_{fi}(\mathbf{R})$ for an electron to make a second-order transition from an initial state $|i\rangle$ to a final state $|f\rangle$ is given by

$$P_{fi}(\mathbf{R}) = \frac{1}{\hbar^4} \int_{-\infty}^{\infty} dt_1' \int_{-\infty}^{\infty} dt_2' \int_{-\infty}^{\infty} dt_1 \int_{-\infty}^{\infty} dt_2 \left(\mathcal{L}_{fi}^{\alpha\beta}(t_1', t_2')\right)^* \mathcal{L}_{fi}^{\gamma\delta}(t_1, t_2) \mathcal{E}_{\beta\alpha\gamma\delta}(\mathbf{R}; t_1', t_2'; t_1, t_2), \qquad (S15)$$

with the second-order dipole correlation function of the electron system at the vacuum-gold interface $\mathcal{L}_{fi}^{\alpha\beta}(t_1, t_2)$ and the second-order field correlation function $\mathcal{E}_{\beta\alpha\gamma\delta}(\mathbf{R}; t_1', t_2'; t_1, t_2)$. Note, that we employed the Einstein summation convention and will do so for the rest of the text and that $\alpha, \beta, \gamma, \delta \in \{x, y, z\}$. We only consider absorption and neglect emission of electromagnetic field quanta. In this case the field correlation function is given by

$$\mathcal{E}_{\beta\alpha\gamma\delta}(\mathbf{R}; t_1', t_2'; t_1, t_2) = \left\langle \hat{E}_\beta^{(-)}(\mathbf{R}, t_2') \hat{E}_\alpha^{(-)}(\mathbf{R}, t_1') \hat{E}_\gamma^{(+)}(\mathbf{R}, t_1) \hat{E}_\delta^{(+)}(\mathbf{R}, t_2) \right\rangle_{\hat{\sigma}}, \qquad (S16)$$

where $\hat{E}_\alpha^{(\pm)}$ is the positive/negative ($\pm$) frequency component of the $\alpha$-th vectorial component of the operator for the quantized electromagnetic field and $\langle \ldots \rangle_{\hat{\sigma}} = \text{Tr}(\hat{\sigma} \ldots)$ is the expectation value with respect to the initial state density matrix $\hat{\sigma}$ of the electric field. The dipole correlation function is given by

$$\mathcal{L}_{fi}^{\alpha\beta}(t_1, t_2) = \Theta(t_1 - t_2) \sum_m (\boldsymbol{\mu}_{fm})^\alpha (\boldsymbol{\mu}_{mi})^\beta e^{i\omega_f t_1} e^{-i\omega_i t_2} e^{-i\omega_m(t_1 - t_2)}, \qquad (S17)$$

where $\boldsymbol{\mu}_{ji} = \langle j|\hat{\mu}|i\rangle$ is the transition dipole matrix-element for the transition $i \to j$, and $\omega_f$, $\omega_m$ and $\omega_i$ are the frequencies of the final, intermediate and initial electronic states, respectively. The sum in Eq. (S17) goes over all possible intermediate states. Causality of the step-wise excitation from the initial to the intermediate state and from the intermediate to the final state is ensured by the Heaviside step function $\Theta(t_1 - t_2)$.

### 1.3.1 Initial and final state average

In a photoemission experiment one does not have direct experimental access to the initial state of the liberated electron. In principle one only can measure the quantum numbers of the final state of the liberated electron, namely its momentum, its kinetic energy and its spin. For a monochromatic light source and simple materials there is a one-to-one correspondence between the initial state energy and the final state kinetic energy of the electron. However, if the light source is not monochromatic, as it is the case for ultrashort laser pulses, and the accessible initial states form a continuum, as it is the case for metals like gold, a magnitude of initial states contributes to a single final state energy [8]. In this case, one has to integrate the emission probability over all initial states.

In photoemission microscopy of SPPs one often does not even do final state resolved imaging, but rather integrates over the final state kinetic energy, momentum and spin. As a result, one only measures the spatially-resolved integral photoemission yield. It is thus natural to average Eq. (S15) over all inital and final states to acquire the integral electron emission probability

$$P(\mathbf{R}) = \frac{1}{\hbar^4} \int_{-\infty}^{\infty} \mathrm{d}t'_1 \int_{-\infty}^{\infty} \mathrm{d}t'_2 \int_{-\infty}^{\infty} \mathrm{d}t_1 \int_{-\infty}^{\infty} \mathrm{d}t_2 \, \mathcal{M}^{\alpha\beta\gamma\delta}(t'_1, t'_2; t_1, t_2) \mathcal{E}_{\beta\alpha\gamma\delta}(\mathbf{R}; t'_1, t'_2; t_1, t_2), \quad (S18)$$

to describe our experiment. We introduced the integral electronic dipole response function

$$\mathcal{M}^{\alpha\beta\gamma\delta}(t'_1, t'_2; t_1, t_2) = \sum_{i,f} \rho_i \left( \mathcal{L}_{fi}^{\alpha\beta}(t'_1, t'_2) \right)^* \mathcal{L}_{fi}^{\gamma\delta}(t_1, t_2), \quad (S19)$$

where $\rho_i = \langle i|\hat{\rho}|i \rangle$ are the diagonal matrix elements of the initial state density matrix of the electron system.

## 1.4 Connection between the quantum-mechanical and the semi-classical model

We know that Eq. (S8) is well-working approximation to the second-order electron yield that is routinely used to describe time-resolved PEEM experiments. Equation (S18) has two major qualitative deviations from the simple form of Eq. (S8). First, the dipole correlation function and the field correlation function both appear at four different times in a convolutional manner. Second, each vectorial component of the electric field in the field correlation function is weighted by different components of the dipole correlation function and furthermore, all possible combinations of field components are included.

In the following we will introduce a small set of approximations, which enable us to resolve these apparent complications and derive a direct quantum-mechanical generalization of Eq. (S9). We will first partially eliminate the convolution integrals in Eq. (S18) by considering the (virtual) state lifetime of the intermediate state in the electron emission process. Then, we will eliminate the directional dependence introduced by the polarization mixing due to the tensorial nature of the electronic dipole response function in Eq. (S18). Finally, we will eliminate the remaining temporal convolution by considering the bandwidth of available second-order electron transitions. Note, that the order in which we introduce the approximations is physically not important but simplifies the following derivations.

### 1.4.1 Short-lived intermediate states

The temporal convolutions in Eq. (S18) would imply that if the dipole correlation function includes long-lived coherences between different times, any temporal cross-correlation between different fields would appear significantly broadened in the time-resolved second-order electron yield. In an extreme case this would imply the liberation of electrons even if the pulses in the field correlation function are well separated in time. This, however, contradicts the experience that the temporal shape of the time-resolved second-order electron yield is typically dominated by the cross-correlation of the probe pulse and the SPP pulse at the surface and usually no severe broadening of the signal is observed.

Partially, the convolutions in Eq. (S18) appear, because the intermediate state of the electron during the second-order emission process is assumed the be a real and infinitely long-lived state. In a real metal, however, electrons have an intrinsic lifetime as they interact with each other via Coulomb scattering and with their crystalline environment via electron-phonon-interaction. If these interactions are weak enough, excitations in the interacting electron system can be well approximated by single-particle excitations of weakly interacting quasi-electrons with a renormalized mass, a renormalized energy and a finite lifetime. This concept is one of the key-ideas behind Fermi liquid theory and especially the finite lifetime of the quasi-electrons is important to describe the temporal evolution of the electron yield in our experiment. To include such effects, we first formally identify the free retarded single-particle Green's function, i.e. the Schrödinger propagator, of the intermediate state within the dipole correlation function Eq. (S17)

$$G^{\mathrm{R}}_{mm'}(t_1 - t_2) = -i\Theta(t_1 - t_2)e^{-i\omega_m(t_1-t_2)}\delta_{mm'}, \tag{S20}$$

such that the dipole correlation function becomes

$$\mathcal{L}^{\alpha\beta}_{fi}(t_1, t_2) = i\sum_m (\boldsymbol{\mu}_{fm})^\alpha (\boldsymbol{\mu}_{mi})^\beta G^{\mathrm{R}}_{mm}(t_1 - t_2)e^{i\omega_f t_1}e^{-i\omega_i t_2}. \tag{S21}$$

As a simplistic approximation, we can now generalize the original theory of two-photon absorption by Mollow [7] to realistic electron systems by replacing the free Green's function with a Green's function describing the quasi-electrons in the metal. A simple ansatz for the quasi-electron Green's function, which includes all essential concepts of weakly interacting quasi-electrons, is given by

$$G^{\mathrm{R}}_{mm'}(t_1 - t_2) = -i\Theta(t_1 - t_2)Z_m e^{-i\omega_m(t_1-t_2)}e^{-(t_1-t_2)/\tau_m}\delta_{mm'}, \tag{S22}$$

where $Z_m$ it the spectral weight and $\tau_m$ is the lifetime of the quasi-electron in the intermediate state $|m\rangle$. The quasi-electron Green's function can be interpreted as the probability amplitude of finding a quasi-electron in the single-particle intermediate state $|m'\rangle$ at time $t_1$ after it was put in the state $|m\rangle$ at time $t_2$. Apparently, this probability amplitude is diagonal in the quantum numbers $m, m'$ and decays exponentially with the lifetime $\tau_m$ as the time-constant.

We now assume that the intermediate state lifetime is sufficiently short compared to the overall dynamics of the field correlation function. This is a good approximation for dominantly virtual intermediate states and especially for the rapid dephasing and decay of optically excited electrons in noble metals like gold [9], [10]. Even if the intermediate state lifetime would be comparable to the dynamics of the field correlation function, broadening effects would typically be weak and could only be found in the temporal wings of second-order electron yield. Under this assumption the exponential decay in the Green's function in Eq. (S22) can be approximated by a $\delta$-function and we have

$$G^{\mathrm{R}}_{mm'}(t_1 - t_2) \approx -i\delta(t_1 - t_2)\tau_m Z_m \delta_{mm'}. \tag{S23}$$

Note that the decay constant appears as a multiplicative factor because of the scaling property of the $\delta$-function. In this approximation the dipole correlation function becomes

$$\begin{aligned}\mathcal{L}^{\gamma\delta}_{fi}(t_1, t_2) &\approx \delta(t_1 - t_2)\mathcal{L}^{\gamma\delta}_{fi}(t_1) \\ &= \delta(t_1 - t_2)e^{i(\omega_f - \omega_i)t_1}\sum_m (\boldsymbol{\mu}_{fm})^\alpha (\boldsymbol{\mu}_{mi})^\beta Z_m \tau_m\end{aligned} \tag{S24}$$

and we can write for the dipole response function

$$\mathcal{M}^{\alpha\beta\gamma\delta}(t'_1, t'_2; t_1, t_2) \approx \delta(t_1 - t_2)\delta(t'_1 - t'_2)\mathcal{M}^{\alpha\beta\gamma\delta}(t_1 - t'_1), \tag{S25}$$

with the integral dipole response function

$$\begin{aligned}\mathcal{M}^{\alpha\beta\gamma\delta}(t_1 - t'_1) &= \sum_{i,f} \rho_i \left(\mathcal{L}^{\alpha\beta}_{fi}(t'_1)\right)^* \mathcal{L}^{\gamma\delta}_{fi}(t_1) \\ &= \sum_{i,f} \rho_i e^{i(\omega_f - \omega_i)(t_1 - t'_1)} \sum_{n,m} Z_n Z_m \tau_n \tau_m (\boldsymbol{\mu}^*_{fn})^\alpha (\boldsymbol{\mu}^*_{ni})^\beta (\boldsymbol{\mu}_{fm})^\gamma (\boldsymbol{\mu}_{mi})^\delta.\end{aligned} \tag{S26}$$

This function only depends on the time difference $\tau = t_1 - t'_1$. If we insert Eq. (S26) into Eq. (S18), integrate out the $\delta$-functions and do a change of variables, we arrive at the expression

$$P(\mathbf{R}) = \frac{1}{\hbar^4} \int_{-\infty}^{\infty} dt \int_{-\infty}^{\infty} d\tau\, \mathcal{M}^{\alpha\beta\gamma\delta}(\tau) \mathcal{E}_{\beta\alpha\gamma\delta}(\mathbf{R}; t, t; t + \tau, t + \tau) \tag{S27}$$

for the electron emission probability. In this form, the field correlation function is only evaluated at two different times, which are connected in a convolutional manner by the material response function.

### 1.4.2 Negligible intermediate state interference

For broadband femtosecond laser pulses multiple intermediate states can fulfill energy conservation for a single pair of initial and final states. This corresponds to the sum over all combinations of intermediate states in Eq. (S26). In time-resolved photoemission one often assumes that different intermediate states, that connect the same initial and final state, do not interfere. Then, one can replace the coherent sum over all combinations of intermediate states by an incoherent sum, such that

$$\mathcal{M}^{\alpha\beta\gamma\delta}(\tau) \approx \sum_{i,f} \rho_i e^{i(\omega_f - \omega_i)\tau} \sum_m Z_m^2 \tau_m^2 (\boldsymbol{\mu}^*_{fm})^\alpha (\boldsymbol{\mu}^*_{mi})^\beta (\boldsymbol{\mu}_{fm})^\gamma (\boldsymbol{\mu}_{mi})^\delta. \tag{S28}$$

### 1.4.3 Isotropic distribution of dipole moment orientations

Usually, the noble metals are considered to be simple metals, i.e. their electronic properties are assumed to be well approximated in a free electron gas model and to have no directional dependence with respect to the crystal lattice. Experimentally there is no strong evidence that on cesiated Au(111) surfaces like the one in our experiment mixing of different field components during the photoemission process or polarization-direction dependencies of the photoemission process dominate the electron yield [11]. Just recently the polarization-isotropy of the emission process even enabled a retrieval of the local electric field polarization at the surface only from the electron yield [12]. We thus consider the transition dipole moments in the metal to be isotropically distributed.

For that reason, we introduce the unit vectors $\boldsymbol{u}_{fm} = \boldsymbol{\mu}_{fm}/|\boldsymbol{\mu}_{fm}|$ and $\boldsymbol{u}_{mi} = \boldsymbol{\mu}_{mi}/|\boldsymbol{\mu}_{mi}|$ and we write for the material response function

$$\mathcal{M}^{\alpha\beta\gamma\delta}(\tau) = \sum_{i,f} \rho_i e^{i(\omega_f - \omega_i)\tau} \sum_m Z_m^2 \tau_m^2 |\boldsymbol{\mu}_{fm}|^2 |\boldsymbol{\mu}_{mi}|^2 (\boldsymbol{u}_{fm}^*)^\alpha (\boldsymbol{u}_{mi}^*)^\beta (\boldsymbol{u}_{fm})^\gamma (\boldsymbol{u}_{mi})^\delta. \tag{S29}$$

Note, that the transition dipole moments generally are complex three-dimensional vectors and thus the introduced unit vectors are also complex three-dimensional vectors. Strictly speaking it does not make much sense to speak of an orientation of the complex unit vectors because of their complex-valued components. Still, we can formulate a notation of an isotropic distribution of transition dipole moments. For this we consider the unit vectors to be random variables which are statistically independent of the modulus squares of the transition dipole moments and all other quantities in Eq. (S29). We then approximate the material response function by its directional average

$$\begin{aligned}\mathcal{M}^{\alpha\beta\gamma\delta}(\tau) &\approx \mathbb{E}\left[\mathcal{M}^{\alpha\beta\gamma\delta}(\tau)\right] \\ &= \sum_{i,f} \rho_i e^{i(\omega_f-\omega_i)\tau} \sum_m Z_m^2 \tau_m^2 |\boldsymbol{\mu}_{fm}|^2 |\boldsymbol{\mu}_{mi}|^2 \mathbb{E}[(\boldsymbol{u}_{fm}^*)^\alpha (\boldsymbol{u}_{fm})^\gamma] \mathbb{E}[(\boldsymbol{u}_{mi}^*)^\beta (\boldsymbol{u}_{mi})^\delta], \end{aligned} \tag{S30}$$

where $\mathbb{E}[\ldots]$ denotes the statistical expectation value.

To model the isotropicity of the dipole response of the metal we consider the unit vectors of the transition dipole moments to be isotropically distributed over the three-dimensional complex unit-sphere. For real vectors such a distribution would be achieved by sampling the components $u_i$ of a vector $\mathbf{u} = (u_x, u_y, u_z)^\mathrm{T}$ from a standard normal distribution $\mathcal{N}(0,1)$ and subsequently normalizing the vectors. That is, if we have $u_i \sim \mathcal{N}(0,1)$ and $u_i$ and $u_j$ independent and identically distributed ($u_i \perp\!\!\!\perp u_j$), then $\mathbf{u}/|\mathbf{u}|$ is istoropically distributed over the three-dimensional unit sphere [13]. For complex vectors, this is generalized by sampling the vector components from a complex normal distribution, i.e. $u_i \sim \mathcal{CN}(0,1)$ and thus $\mathrm{Re}\{u_i\} \sim \mathcal{N}(0,1/2)$, $\mathrm{Im}\{u_i\} \sim \mathcal{N}(0,1/2)$ and $\mathrm{Re}\{u_i\} \perp\!\!\!\perp \mathrm{Im}\{u_i\}$.

The probability density function $f(\mathbf{u})$ for the three-dimensional complex case is given by

$$f(\mathbf{u}) = \frac{1}{\pi^3} \exp\left(-\sum_{i=(x,y,z)} |u_i|^2\right) = \frac{1}{\pi^3} \exp\left(-|\mathbf{u}|^2\right). \tag{S31}$$

It is obvious, that this probability density function is invariant under unitary transformations, which directly reflects the isotropicity of the distribution. In the following we will derive expressions for the expectation values $\mathbb{E}[(\boldsymbol{u}_{fm}^*)^\alpha(\boldsymbol{u}_{fm})^\gamma]$ and $\mathbb{E}[(\boldsymbol{u}_{mi}^*)^\delta(\boldsymbol{u}_{mi})^\beta]$ based on this probability distribution function. Per construction we have $|\mathbf{u}/|\mathbf{u}||^2 = 1$ and it follows that

$$1 = \mathbb{E}\left[\sum_{i=(x,y,z)}\left(\frac{|u_i|}{|\mathbf{u}|}\right)^2\right] = 3\mathbb{E}\left[\frac{|u_i|^2}{|\mathbf{u}|^2}\right] \Leftrightarrow \mathbb{E}\left[\frac{|u_i|^2}{|\mathbf{u}|^2}\right] = \frac{1}{3}, \tag{S32}$$

from which we conclude that $\mathbb{E}[(\boldsymbol{u}_{fm}^*)^\alpha(\boldsymbol{u}_{fm})^\alpha] = 1/3$ and $\mathbb{E}[(\boldsymbol{u}_{mi}^*)^\delta(\boldsymbol{u}_{mi})^\delta] = 1/3$. We are left with calculating the off-diagonal expectation values, i.e. expectation values of the form $\mathbb{E}[u_i u_j^*/|\mathbf{u}|^2]$. From the unitary symmetry of the probability density function Eq. (S31) it follows that these expectation values need to be invariant under the change $u_i \to u_i e^{i\phi}$, i.e.

$$\mathbb{E}[u_i u_j^*/|\mathbf{u}|^2] = \mathbb{E}[u_i e^{i\phi} u_j^*/|\mathbf{u}|^2] = e^{i\phi}\mathbb{E}[u_i u_j^*/|\mathbf{u}|^2]. \tag{S33}$$

As this equality needs to hold for any $\phi \in \mathbb{R}$, we conclude that the off-diagonal expectation values need to vanish and we finally have $\mathbb{E}[(\boldsymbol{u}_{fm}^*)^\alpha(\boldsymbol{u}_{fm})^\gamma] = \delta^{\alpha\gamma}/3$ and $\mathbb{E}[(\boldsymbol{u}_{mi}^*)^\beta(\boldsymbol{u}_{mi})^\delta] = \delta^{\beta\delta}/3$. Combining this result and Eq. (S30) we finally arrive at

$$\mathcal{M}^{\alpha\beta\gamma\delta}(\tau) \approx \frac{1}{9}\delta^{\alpha\gamma}\delta^{\beta\delta}\mathcal{M}(\tau) = \frac{1}{9}\delta^{\alpha\gamma}\delta^{\beta\delta}\sum_{i,f}\rho_i e^{i(\omega_f-\omega_i)\tau}\sum_m Z_m^2\tau_m^2|\boldsymbol{\mu}_{fm}|^2|\boldsymbol{\mu}_{mi}|^2, \tag{S34}$$

where we introduced the directionally independent material response function

$$\mathcal{M}(\tau) = \sum_{i,f}\rho_i e^{i(\omega_f-\omega_i)\tau}\sum_m Z_m^2\tau_m^2|\boldsymbol{\mu}_{fm}|^2|\boldsymbol{\mu}_{mi}|^2. \tag{S35}$$

In this approximate the integral electron emission probability in Eq. (S27) becomes

$$\begin{aligned}P(\mathbf{R}) &\approx \frac{1}{9\hbar^4}\int_{-\infty}^{\infty}dt\int_{-\infty}^{\infty}d\tau\,\mathcal{M}(\tau)\delta^{\alpha\gamma}\delta^{\beta\delta}\mathcal{E}_{\beta\alpha\gamma\delta}(\mathbf{R};t,t;t+\tau,t+\tau) \\ &= \frac{1}{9\hbar^4}\int_{-\infty}^{\infty}dt\int_{-\infty}^{\infty}d\tau\,\mathcal{M}(\tau)\mathcal{E}_{\beta\alpha}{}^{\alpha\beta}(\mathbf{R};t,t;t+\tau,t+\tau).\end{aligned} \tag{S36}$$

### 1.4.4 Instantaneous material response

The laser pulses used in the experiment typically have a bandwidth of $\sim 80\,\text{nm}$ centered around $800\,\text{nm}$, or in terms of energies a bandwidth of $\sim 155\,\text{meV}$ centered around $1.55\,\text{eV}$. This bandwidth is considerably small compared to the bandwidth of available two-photon transitions for the cesiated Au(111) surface. Any coherences between different second-order transitions within this bandwidth into the continuum of free electron final states are expected to average out in the integral electron emission probability [8]. To include this, we first notice that the material response function in Eq (S35) is nothing else than a Fourier sum, which might be rewritten as

$$\mathcal{M}(\tau) = \frac{1}{2\pi}\sum_{i,f}\mathcal{M}_{fi}e^{i\omega_{fi}\tau} \tag{S37}$$

with the matrix elements $\mathcal{M}_{fi} = 2\pi \rho_i \sum_m Z_m^2 \tau_m^2 |\boldsymbol{\mu}_{fm}|^2 |\boldsymbol{\mu}_{mi}|^2$ and the transition frequency $\omega_{fi} = \omega_f - \omega_i$. Then, the integral electron emission probability in Eq. (S36) can be written as

$$P(\mathbf{R}) = \frac{1}{2\pi} \frac{1}{9\hbar^4} \sum_{i,f} \mathcal{M}_{fi} \int_{-\infty}^{\infty} \mathrm{d}t \int_{-\infty}^{\infty} \mathrm{d}\tau \, e^{i\omega_{fi}\tau} \mathcal{E}_{\beta\alpha}{}^{\alpha\beta}\left(\mathbf{R}; t, t; t+\tau, t+\tau\right). \tag{S38}$$

We now switch to frequency space using the frequency representation of the positive and negative frequency parts of the electric field operators in the field correlation function Eq. (S16)

$$\hat{E}_\alpha^{(\pm)}(\mathbf{R}, t) = \frac{1}{2\pi} \int_{-\infty}^{\infty} \mathrm{d}t \, e^{\mp i\omega t} \hat{E}_\alpha^{(\pm)}(\mathbf{R}, \omega). \tag{S39}$$

By doing so and integrating out any appearing delta functions we arrive at the integral electron emission probability

$$P(\mathbf{R}) = \frac{1}{2\pi} \frac{1}{9\hbar^4} \sum_{i,f} \mathcal{M}_{fi} \int_{-\infty}^{\infty} \frac{\mathrm{d}\omega}{2\pi} \int_{-\infty}^{\infty} \frac{\mathrm{d}\omega'}{2\pi} \tilde{\mathcal{E}}_{\beta\alpha}{}^{\alpha\beta}\left(\mathbf{R}; \omega_{fi} - \omega', \omega'; \omega_{fi} - \omega, \omega\right) \tag{S40}$$

with the frequency representation of the field correlation function $\tilde{\mathcal{E}}_{\beta\alpha}{}^{\alpha\beta}(\mathbf{R}; \omega_1', \omega_2'; \omega_1, \omega_2)$. Now it becomes apparent that the matrix elements $\mathcal{M}_{fi}$ just act as weights for the the different frequency components of the field correlation function. If the matrix elements are sufficiently constant over the bandwidth of the field correlation function we might assume $\mathcal{M}_{fi} \approx \mathcal{M}$ and thus write for the integral electron emission probability in its frequency representation

$$P(\mathbf{R}) \approx \frac{1}{2\pi} \frac{\mathcal{M}}{9\hbar^4} \sum_{i,f} \int_{-\infty}^{\infty} \frac{\mathrm{d}\omega}{2\pi} \int_{-\infty}^{\infty} \frac{\mathrm{d}\omega'}{2\pi} \tilde{\mathcal{E}}_{\beta\alpha}{}^{\alpha\beta}\left(\mathbf{R}; \omega_{fi} - \omega', \omega'; \omega_{fi} - \omega, \omega\right). \tag{S41}$$

Switching back to the time-domain leaves us with

$$P(\mathbf{R}) = \frac{1}{2\pi} \frac{\mathcal{M}}{9\hbar^4} \sum_{i,f} \int_{-\infty}^{\infty} \mathrm{d}t' \int_{-\infty}^{\infty} \mathrm{d}t \, e^{i\omega_{fi}(t-t')} \mathcal{E}_{\beta\alpha}{}^{\alpha\beta}\left(\mathbf{R}; t', t'; t, t\right). \tag{S42}$$

If the bandwidth of available second-order transitions $i \to f$ is large compared to the bandwidth of the field correlation function we might switch to the appropriate continuum limit. If we introduce the transition density $\rho(\omega) = \sum_{i,f} \delta(\omega - \omega_{fi})$, we can write

$$P(\mathbf{R}) = \frac{1}{2\pi} \frac{\mathcal{M}}{9\hbar^4} \int_{-\infty}^{\infty} \mathrm{d}t' \int_{-\infty}^{\infty} \mathrm{d}t \int_{-\infty}^{\infty} \mathrm{d}\omega \, \rho(\omega) e^{i\omega(t-t')} \mathcal{E}_{\beta\alpha}{}^{\alpha\beta}\left(\mathbf{R}; t', t'; t, t\right). \tag{S43}$$

Assuming that the transition density is sufficiently broadband compared to the bandwidth of the field correlation function, it is reasonable to approximate $\int_{-\infty}^{\infty} \mathrm{d}\omega \, \rho(\omega) e^{i\omega(t-t')} \approx 2\pi \delta(t-t')$ and we finally arrive (after integrating out this delta function) at the integral electron emission probability

$$P(\mathbf{R}) = \frac{\mathcal{M}}{9\hbar^4} \int_{-\infty}^{\infty} \mathrm{d}t \, \mathcal{E}_{\beta\alpha}{}^{\alpha\beta}\left(\mathbf{R}; t, t; t, t\right). \tag{S44}$$

We note that this equation is a direct quantum mechanical generalization of Eq. (S8), where all material degrees of freedom have been integrated out.

## 1.5 Field correlation function

We are left with computing the equal-time field correlation function. We first explicitly split the electric field operator into the SPP mode and the probe field mode $\hat{\mathbf{E}}^{(\pm)}(\mathbf{R},t) = \hat{\mathbf{E}}^{(\pm)}_{\text{SPP}}(\mathbf{R},t) + \hat{\mathbf{E}}^{(\pm)}_{\text{Probe}}(\mathbf{R},t-\Delta t)$, where $\Delta t$ is the pump-probe delay. Then, the field correlation function can be expanded as

$$\begin{aligned}
\mathcal{E}_{\beta\alpha}{}^{\alpha\beta}(\mathbf{R},\Delta t) =& \left\langle \left(\hat{\mathbf{E}}^{(-)}_{\text{Probe}}\right)_\beta \left(\hat{\mathbf{E}}^{(-)}_{\text{Probe}}\right)_\alpha \left(\hat{\mathbf{E}}^{(+)}_{\text{Probe}}\right)^\alpha \left(\hat{\mathbf{E}}^{(+)}_{\text{Probe}}\right)^\beta \right\rangle_{\hat{\sigma}} \\
&+ 2\,\text{Re}\left\{ \left\langle \left(\hat{\mathbf{E}}^{(-)}_{\text{Probe}}\right)_\beta \left(\hat{\mathbf{E}}^{(-)}_{\text{Probe}}\right)_\alpha \left(\hat{\mathbf{E}}^{(+)}_{\text{Probe}}\right)^\alpha \left(\hat{\mathbf{E}}^{(+)}_{\text{SPP}}\right)^\beta \right\rangle_{\hat{\sigma}} \right\} \\
&+ 2\,\text{Re}\left\{ \left\langle \left(\hat{\mathbf{E}}^{(-)}_{\text{Probe}}\right)_\beta \left(\hat{\mathbf{E}}^{(-)}_{\text{Probe}}\right)_\alpha \left(\hat{\mathbf{E}}^{(+)}_{\text{SPP}}\right)^\alpha \left(\hat{\mathbf{E}}^{(+)}_{\text{Probe}}\right)^\beta \right\rangle_{\hat{\sigma}} \right\} \\
&+ 2\,\text{Re}\left\{ \left\langle \left(\hat{\mathbf{E}}^{(-)}_{\text{Probe}}\right)_\beta \left(\hat{\mathbf{E}}^{(-)}_{\text{Probe}}\right)_\alpha \left(\hat{\mathbf{E}}^{(+)}_{\text{SPP}}\right)^\alpha \left(\hat{\mathbf{E}}^{(+)}_{\text{SPP}}\right)^\beta \right\rangle_{\hat{\sigma}} \right\} \\
&+ 2\,\text{Re}\left\{ \left\langle \left(\hat{\mathbf{E}}^{(-)}_{\text{Probe}}\right)_\beta \left(\hat{\mathbf{E}}^{(-)}_{\text{SPP}}\right)_\alpha \left(\hat{\mathbf{E}}^{(+)}_{\text{Probe}}\right)^\alpha \left(\hat{\mathbf{E}}^{(+)}_{\text{SPP}}\right)^\beta \right\rangle_{\hat{\sigma}} \right\} \\
&+ \left\langle \left(\hat{\mathbf{E}}^{(-)}_{\text{SPP}}\right)_\beta \left(\hat{\mathbf{E}}^{(-)}_{\text{Probe}}\right)_\alpha \left(\hat{\mathbf{E}}^{(+)}_{\text{Probe}}\right)^\alpha \left(\hat{\mathbf{E}}^{(+)}_{\text{SPP}}\right)^\beta \right\rangle_{\hat{\sigma}} \\
&+ \left\langle \left(\hat{\mathbf{E}}^{(-)}_{\text{Probe}}\right)_\beta \left(\hat{\mathbf{E}}^{(-)}_{\text{SPP}}\right)_\alpha \left(\hat{\mathbf{E}}^{(+)}_{\text{SPP}}\right)^\alpha \left(\hat{\mathbf{E}}^{(+)}_{\text{Probe}}\right)^\beta \right\rangle_{\hat{\sigma}} \\
&+ 2\,\text{Re}\left\{ \left\langle \left(\hat{\mathbf{E}}^{(-)}_{\text{SPP}}\right)_\beta \left(\hat{\mathbf{E}}^{(-)}_{\text{SPP}}\right)_\alpha \left(\hat{\mathbf{E}}^{(+)}_{\text{SPP}}\right)^\alpha \left(\hat{\mathbf{E}}^{(+)}_{\text{Probe}}\right)^\beta \right\rangle_{\hat{\sigma}} \right\} \\
&+ 2\,\text{Re}\left\{ \left\langle \left(\hat{\mathbf{E}}^{(-)}_{\text{SPP}}\right)_\beta \left(\hat{\mathbf{E}}^{(-)}_{\text{SPP}}\right)_\alpha \left(\hat{\mathbf{E}}^{(+)}_{\text{Probe}}\right)^\alpha \left(\hat{\mathbf{E}}^{(+)}_{\text{SPP}}\right)^\beta \right\rangle_{\hat{\sigma}} \right\} \\
&+ \left\langle \left(\hat{\mathbf{E}}^{(-)}_{\text{SPP}}\right)_\beta \left(\hat{\mathbf{E}}^{(-)}_{\text{SPP}}\right)_\alpha \left(\hat{\mathbf{E}}^{(+)}_{\text{SPP}}\right)^\alpha \left(\hat{\mathbf{E}}^{(+)}_{\text{SPP}}\right)^\beta \right\rangle_{\hat{\sigma}},
\end{aligned} \quad (S45)$$

where we left out the arguments of the field operators for clarity. Clearly, the field correlation function includes all possible fourth-order mixings of the SPP and the probe field operators.

To proceed with the calculation of the field correlation function we first need to quantize the SPP and the probe field as defined in Sec. 1.1. For this we employ standard canonical quantization of the defined mode functions [14] such that we have

$$\hat{\mathbf{E}}^{(+)}_{\text{Probe}}(\mathbf{r},t) = i\sqrt{\frac{\hbar\omega}{2\varepsilon_0 V}}\frac{1}{\sqrt{\mathcal{N}_{\text{Probe}}}}\boldsymbol{\epsilon}_{\text{Probe}} u_{\text{Probe}}(\mathbf{r})e^{-i\omega t}\hat{a}, \quad (S46a)$$

$$\hat{\mathbf{E}}^{(+)}_{\text{SPP}}(\mathbf{r},t) = i\sqrt{\frac{\hbar\omega}{2\varepsilon_0 V}}\frac{1}{\sqrt{\mathcal{N}_{\text{SPP}}}}\boldsymbol{\epsilon}_{\text{SPP}}(z) u_{\text{SPP}}(\mathbf{r})e^{-i\omega t}\hat{b}, \quad (S46b)$$

where we introduced the annihilation operator for the probe field mode $\hat{a}$ and for the SPP mode $\hat{b}$. We also introduced the quantization volume $V$ and appropriate normalization constants $\mathcal{N}_{\text{Probe}}$ and $\mathcal{N}_{\text{SPP}}$ for the mode functions, which's values are not relevant to our discussion. Using the same arguments as in Sec. 1.2 we again set $z = 0^-$ and thus approximate the the field operators by

$$\hat{\mathbf{E}}_{\text{Probe}}^{(+)}(\mathbf{R}, t) = i\sqrt{\frac{\hbar\omega}{2\varepsilon_0 V}} \frac{1}{\sqrt{\mathcal{N}_{\text{Probe}}}} \boldsymbol{\epsilon}_{\text{Probe}} e^{-i\omega t} \hat{a} \tag{S47a}$$

$$\hat{\mathbf{E}}_{\text{SPP}}^{(+)}(\mathbf{R}, t) = i\sqrt{\frac{\hbar\omega}{2\varepsilon_0 V}} \frac{1}{\sqrt{\mathcal{N}_{\text{SPP}}}} \boldsymbol{\epsilon}_{\text{SPP}} e^{i\mathbf{k}_{\text{SPP}} \cdot \mathbf{r}} e^{-i\omega t} \hat{b}, \tag{S47b}$$

where we absorbed the amplitude transmission coefficient into the normalization constant of the probe field $\mathcal{N}_{\text{Probe}} \to \mathcal{N}_{\text{Probe}} \mathcal{T}^2$.

### 1.5.1 Transition Rate

We will see in the following section that the explicit time-dependence of the field correlation function cancels out and we are only left with a dependence on the pump-probe delay $\Delta t$, i.e. the time between the excitation of the SPP mode and the arrival of the probe laser pulse at the sample surface. Just like in Sec. 1.2.1 we thus switch to the electron emission probability per unit-time, i.e. the transition rate

$$\Gamma_{\text{2PPE}}^{\text{quantum}}(\mathbf{R}) = \frac{dP(\mathbf{R})}{dt} = \frac{\mathcal{M}}{9\hbar^4} \lim_{t \to \infty} \frac{d}{dt} \int_{-\infty}^{t} dt' \, \mathcal{E}_{\beta\alpha}{}^{\alpha\beta}(\mathbf{R}; t', t'; t', t') = \frac{\mathcal{M}}{9\hbar^4} \mathcal{E}_{\beta\alpha}{}^{\alpha\beta}(\mathbf{R}, \Delta t), \tag{S48}$$

which is a direct quantum mechanical generalization of Eq. (S9). We note that our experiment essentially measures the fourth-order field correlation function of the probe field and the SPP field.

We now insert the electric field operators for the SPP field and the probe field Eq. (S47) into the field correlation function Eq. (S45) and arrive at the transition rate

$$\begin{aligned}
\Gamma_{\text{2PPE}}^{\text{quantum}}(\mathbf{R}, \Delta t) =\ & \frac{\mathcal{M}}{9\hbar^4} \left(\frac{\hbar\omega}{2\varepsilon_0 V}\right)^2 \frac{|\boldsymbol{\epsilon}_{\text{Probe}}|^4}{\mathcal{N}_{\text{Probe}}^2} \langle \hat{a}^\dagger \hat{a}^\dagger \hat{a} \hat{a} \rangle_{\hat{\sigma}} + \frac{\mathcal{M}}{9\hbar^4} \left(\frac{\hbar\omega}{2\varepsilon_0 V}\right)^2 \frac{|\boldsymbol{\epsilon}_{\text{SPP}}|^4}{\mathcal{N}_{\text{SPP}}^2} \langle \hat{b}^\dagger \hat{b}^\dagger \hat{b} \hat{b} \rangle_{\hat{\sigma}} \\
& + \frac{\mathcal{M}}{9\hbar^4} \left(\frac{\hbar\omega}{2\varepsilon_0 V}\right)^2 \frac{|\boldsymbol{\epsilon}_{\text{Probe}}|^2 |\boldsymbol{\epsilon}_{\text{SPP}}|^2}{\mathcal{N}_{\text{Probe}} \mathcal{N}_{\text{SPP}}} \\
& \times \left( \langle \hat{b}^\dagger \hat{a}^\dagger \hat{a} \hat{b} \rangle_{\hat{\sigma}} + \langle \hat{a}^\dagger \hat{b}^\dagger \hat{b} \hat{a} \rangle_{\hat{\sigma}} + \frac{|\boldsymbol{\epsilon}_{\text{Probe}}^* \cdot \boldsymbol{\epsilon}_{\text{SPP}}|^2}{|\boldsymbol{\epsilon}_{\text{Probe}}|^2 |\boldsymbol{\epsilon}_{\text{SPP}}|^2} \left( \langle \hat{a}^\dagger \hat{b}^\dagger \hat{a} \hat{b} \rangle_{\hat{\sigma}} + \langle \hat{b}^\dagger \hat{a}^\dagger \hat{b} \hat{a} \rangle_{\hat{\sigma}} \right) \right) \\
& + 2 \frac{\mathcal{M}}{9\hbar^4} \left(\frac{\hbar\omega}{2\varepsilon_0 V}\right)^2 \text{Re} \left\{ \frac{\boldsymbol{\epsilon}_{\text{Probe}}^* \cdot \boldsymbol{\epsilon}_{\text{SPP}}}{\sqrt{\mathcal{N}_{\text{Probe}}} \sqrt{\mathcal{N}_{\text{SPP}}}} e^{i(\mathbf{k}_{\text{SPP}} \cdot \mathbf{R} - \omega \Delta t)} \right. \\
& \times \left. \left( \frac{|\boldsymbol{\epsilon}_{\text{Probe}}|^2}{\mathcal{N}_{\text{Probe}}} \left( \langle \hat{a}^\dagger \hat{a}^\dagger \hat{a} \hat{b} \rangle_{\hat{\sigma}} + \langle \hat{a}^\dagger \hat{a}^\dagger \hat{b} \hat{a} \rangle_{\hat{\sigma}} \right) + \frac{|\boldsymbol{\epsilon}_{\text{SPP}}|^2}{\mathcal{N}_{\text{SPP}}} \left( \langle \hat{a}^\dagger \hat{b}^\dagger \hat{b} \hat{b} \rangle_{\hat{\sigma}} + \langle \hat{b}^\dagger \hat{a}^\dagger \hat{b} \hat{b} \rangle_{\hat{\sigma}} \right) \right) \right\} \\
& + 2 \frac{\mathcal{M}}{9\hbar^4} \left(\frac{\hbar\omega}{2\varepsilon_0 V}\right)^2 \frac{|\boldsymbol{\epsilon}_{\text{Probe}}^* \cdot \boldsymbol{\epsilon}_{\text{SPP}}|^2}{\mathcal{N}_{\text{Probe}} \mathcal{N}_{\text{SPP}}} \text{Re} \left\{ e^{i(2\mathbf{k}_{\text{SPP}} \cdot \mathbf{R} - 2\omega \Delta t)} \langle \hat{a}^\dagger \hat{a}^\dagger \hat{b} \hat{b} \rangle_{\hat{\sigma}} \right\}.
\end{aligned}$$
(S49)

Formally, this transition rate has a similar structure like Eq. (S10), except for the appearance of quantum-mechanical constants and expectation values of fourth-order mixings of the creation and annihilation operators for the fields.

Again, like in our classical treatment, we switch to reciprocal space by a Fourier-transform of Eq. (S49) in the 2-D position coordinate $\mathbf{R}$. We arrive at the transition rate

$$\begin{aligned}
\Gamma_{\text{2PPE}}^{\text{quantum}}(\mathbf{K}, \Delta t) =& \frac{\mathcal{M}}{9\hbar^4}(2\pi)^2\delta(\mathbf{K})\left(\frac{\hbar\omega}{2\varepsilon_0 V}\right)^2 \frac{|\boldsymbol{\epsilon}_{\text{Probe}}|^4}{\mathcal{N}_{\text{Probe}}^2}\langle \hat{a}^\dagger \hat{a}^\dagger \hat{a}\hat{a}\rangle_{\hat{\sigma}} \\
&+ \frac{\mathcal{M}}{9\hbar^4}(2\pi)^2\delta(\mathbf{K})\left(\frac{\hbar\omega}{2\varepsilon_0 V}\right)^2 \frac{|\boldsymbol{\epsilon}_{\text{SPP}}|^4}{\mathcal{N}_{\text{SPP}}^2}\langle \hat{b}^\dagger \hat{b}^\dagger \hat{b}\hat{b}\rangle_{\hat{\sigma}} \\
&+ \frac{\mathcal{M}}{9\hbar^4}(2\pi)^2\delta(\mathbf{K})\left(\frac{\hbar\omega}{2\varepsilon_0 V}\right)^2 \frac{|\boldsymbol{\epsilon}_{\text{Probe}}|^2|\boldsymbol{\epsilon}_{\text{SPP}}|^2}{\mathcal{N}_{\text{Probe}}\mathcal{N}_{\text{SPP}}} \\
&\times \left(\langle \hat{b}^\dagger \hat{a}^\dagger \hat{a}\hat{b}\rangle_{\hat{\sigma}} + \langle \hat{a}^\dagger \hat{b}^\dagger \hat{b}\hat{a}\rangle_{\hat{\sigma}} + \frac{|\boldsymbol{\epsilon}_{\text{Probe}}^* \cdot \boldsymbol{\epsilon}_{\text{SPP}}|^2}{|\boldsymbol{\epsilon}_{\text{Probe}}|^2|\boldsymbol{\epsilon}_{\text{SPP}}|^2}\left(\langle \hat{a}^\dagger \hat{b}^\dagger \hat{a}\hat{b}\rangle_{\hat{\sigma}} + \langle \hat{b}^\dagger \hat{a}^\dagger \hat{b}\hat{a}\rangle_{\hat{\sigma}}\right)\right) \\
&+ \frac{\mathcal{M}}{9\hbar^4}(2\pi)^2\delta(\mathbf{K} - \mathbf{k}_{\text{SPP}})\left(\frac{\hbar\omega}{2\varepsilon_0 V}\right)^2 \frac{\boldsymbol{\epsilon}_{\text{Probe}}^* \cdot \boldsymbol{\epsilon}_{\text{SPP}}}{\sqrt{\mathcal{N}_{\text{Probe}}}\sqrt{\mathcal{N}_{\text{SPP}}}}e^{-i\omega\Delta t} \\
&\times \left(\frac{|\boldsymbol{\epsilon}_{\text{Probe}}|^2}{\mathcal{N}_{\text{Probe}}}\left(\langle \hat{a}^\dagger \hat{a}^\dagger \hat{a}\hat{b}\rangle_{\hat{\sigma}} + \langle \hat{a}^\dagger \hat{a}^\dagger \hat{b}\hat{a}\rangle_{\hat{\sigma}}\right) + \frac{|\boldsymbol{\epsilon}_{\text{SPP}}|^2}{\mathcal{N}_{\text{SPP}}}\left(\langle \hat{a}^\dagger \hat{b}^\dagger \hat{b}\hat{b}\rangle_{\hat{\sigma}} + \langle \hat{b}^\dagger \hat{a}^\dagger \hat{b}\hat{b}\rangle_{\hat{\sigma}}\right)\right) \\
&+ \frac{\mathcal{M}}{9\hbar^4}(2\pi)^2\delta(\mathbf{K} + \mathbf{k}_{\text{SPP}})\left(\frac{\hbar\omega}{2\varepsilon_0 V}\right)^2 \frac{\boldsymbol{\epsilon}_{\text{SPP}}^* \cdot \boldsymbol{\epsilon}_{\text{Probe}}}{\sqrt{\mathcal{N}_{\text{Probe}}}\sqrt{\mathcal{N}_{\text{SPP}}}}e^{i\omega\Delta t} \\
&\times \left(\frac{|\boldsymbol{\epsilon}_{\text{Probe}}|^2}{\mathcal{N}_{\text{Probe}}}\left(\langle \hat{b}^\dagger \hat{a}^\dagger \hat{a}\hat{a}\rangle_{\hat{\sigma}} + \langle \hat{a}^\dagger \hat{b}^\dagger \hat{a}\hat{a}\rangle_{\hat{\sigma}}\right) + \frac{|\boldsymbol{\epsilon}_{\text{SPP}}|^2}{\mathcal{N}_{\text{SPP}}}\left(\langle \hat{b}^\dagger \hat{b}^\dagger \hat{b}\hat{a}\rangle_{\hat{\sigma}} + \langle \hat{b}^\dagger \hat{b}^\dagger \hat{a}\hat{b}\rangle_{\hat{\sigma}}\right)\right) \\
&+ \frac{\mathcal{M}}{9\hbar^4}(2\pi)^2\delta(\mathbf{K} - 2\mathbf{k}_{\text{SPP}})\left(\frac{\hbar\omega}{2\varepsilon_0 V}\right)^2 \frac{(\boldsymbol{\epsilon}_{\text{Probe}}^* \cdot \boldsymbol{\epsilon}_{\text{SPP}})^2}{\mathcal{N}_{\text{Probe}}\mathcal{N}_{\text{SPP}}}e^{-2i\omega\Delta t}\langle \hat{a}^\dagger \hat{a}^\dagger \hat{b}\hat{b}\rangle_{\hat{\sigma}} \\
&+ \frac{\mathcal{M}}{9\hbar^4}(2\pi)^2\delta(\mathbf{K} + 2\mathbf{k}_{\text{SPP}})\left(\frac{\hbar\omega}{2\varepsilon_0 V}\right)^2 \frac{(\boldsymbol{\epsilon}_{\text{SPP}}^* \cdot \boldsymbol{\epsilon}_{\text{Probe}})^2}{\mathcal{N}_{\text{Probe}}\mathcal{N}_{\text{SPP}}}e^{2i\omega\Delta t}\langle \hat{b}^\dagger \hat{b}^\dagger \hat{a}\hat{a}\rangle_{\hat{\sigma}}\,.
\end{aligned}$$
(S50)

In our experiment we have the probe polarization collinear with the in-plane component of the SPP polarization. By considering the orientation of these polarizations we further simplify the transition rate and arrive at

$$\begin{aligned}
\Gamma_{\text{2PPE}}^{\text{quantum}}(\mathbf{K}, \Delta t) =\ & \frac{\mathcal{M}}{9\hbar^4}(2\pi)^2\delta(\mathbf{K})\left(\frac{\hbar\omega}{2\varepsilon_0 V}\right)^2 \frac{\langle \hat{a}^\dagger\hat{a}^\dagger\hat{a}\hat{a}\rangle_{\hat{\sigma}}}{\mathcal{N}_{\text{Probe}}^2} \\
& + \frac{\mathcal{M}}{9\hbar^4}(2\pi)^2\delta(\mathbf{K})\left(\frac{\hbar\omega}{2\varepsilon_0 V}\right)^2 \left(1 + |\boldsymbol{\epsilon}_{\text{SPP}}^\perp|^2\right)^2 \frac{\langle \hat{b}^\dagger\hat{b}^\dagger\hat{b}\hat{b}\rangle_{\hat{\sigma}}}{\mathcal{N}_{\text{SPP}}^2} \\
& + \frac{\mathcal{M}}{9\hbar^4}(2\pi)^2\delta(\mathbf{K})\left(\frac{\hbar\omega}{2\varepsilon_0 V}\right)^2 \frac{1}{\mathcal{N}_{\text{Probe}}\mathcal{N}_{\text{SPP}}} \\
& \times \left(\langle \hat{a}^\dagger\hat{b}^\dagger\hat{a}\hat{b}\rangle_{\hat{\sigma}} + \langle \hat{b}^\dagger\hat{a}^\dagger\hat{b}\hat{a}\rangle_{\hat{\sigma}} + \left(1 + |\boldsymbol{\epsilon}_{\text{SPP}}^\perp|^2\right)\left(\langle \hat{b}^\dagger\hat{a}^\dagger\hat{a}\hat{b}\rangle_{\hat{\sigma}} + \langle \hat{a}^\dagger\hat{b}^\dagger\hat{b}\hat{a}\rangle_{\hat{\sigma}}\right)\right) \\
& + \frac{\mathcal{M}}{9\hbar^4}(2\pi)^2\delta(\mathbf{K} - \mathbf{k}_{\text{SPP}})\left(\frac{\hbar\omega}{2\varepsilon_0 V}\right)^2 \frac{1}{\sqrt{\mathcal{N}_{\text{Probe}}}\sqrt{\mathcal{N}_{\text{SPP}}}} e^{-i\omega\Delta t} \\
& \times \left(\frac{\langle \hat{a}^\dagger\hat{a}^\dagger\hat{a}\hat{b}\rangle_{\hat{\sigma}} + \langle \hat{a}^\dagger\hat{a}^\dagger\hat{b}\hat{a}\rangle_{\hat{\sigma}}}{\mathcal{N}_{\text{Probe}}} + \left(1 + |\boldsymbol{\epsilon}_{\text{SPP}}^\perp|^2\right)\frac{\langle \hat{a}^\dagger\hat{b}^\dagger\hat{b}\hat{b}\rangle_{\hat{\sigma}} + \langle \hat{b}^\dagger\hat{a}^\dagger\hat{b}\hat{b}\rangle_{\hat{\sigma}}}{\mathcal{N}_{\text{SPP}}}\right) \\
& + \frac{\mathcal{M}}{9\hbar^4}(2\pi)^2\delta(\mathbf{K} + \mathbf{k}_{\text{SPP}})\left(\frac{\hbar\omega}{2\varepsilon_0 V}\right)^2 \frac{1}{\sqrt{\mathcal{N}_{\text{Probe}}}\sqrt{\mathcal{N}_{\text{SPP}}}} e^{i\omega\Delta t} \\
& \times \left(\frac{\langle \hat{b}^\dagger\hat{a}^\dagger\hat{a}\hat{a}\rangle_{\hat{\sigma}} + \langle \hat{a}^\dagger\hat{b}^\dagger\hat{a}\hat{a}\rangle_{\hat{\sigma}}}{\mathcal{N}_{\text{Probe}}} + \left(1 + |\boldsymbol{\epsilon}_{\text{SPP}}^\perp|^2\right)\frac{\langle \hat{b}^\dagger\hat{b}^\dagger\hat{b}\hat{a}\rangle_{\hat{\sigma}} + \langle \hat{b}^\dagger\hat{b}^\dagger\hat{a}\hat{b}\rangle_{\hat{\sigma}}}{\mathcal{N}_{\text{SPP}}}\right) \\
& + \frac{\mathcal{M}}{9\hbar^4}(2\pi)^2\delta(\mathbf{K} - 2\mathbf{k}_{\text{SPP}})\left(\frac{\hbar\omega}{2\varepsilon_0 V}\right)^2 e^{-2i\omega\Delta t}\frac{\langle \hat{a}^\dagger\hat{a}^\dagger\hat{b}\hat{b}\rangle_{\hat{\sigma}}}{\mathcal{N}_{\text{Probe}}\mathcal{N}_{\text{SPP}}} \\
& + \frac{\mathcal{M}}{9\hbar^4}(2\pi)^2\delta(\mathbf{K} + 2\mathbf{k}_{\text{SPP}})\left(\frac{\hbar\omega}{2\varepsilon_0 V}\right)^2 e^{2i\omega\Delta t}\frac{\langle \hat{b}^\dagger\hat{b}^\dagger\hat{a}\hat{a}\rangle_{\hat{\sigma}}}{\mathcal{N}_{\text{Probe}}\mathcal{N}_{\text{SPP}}}.
\end{aligned} \quad (S51)$$

### 1.5.2 Connection to experimental intensities

While Eq. (S51) in principle enables a full quantum-mechanical description of the electron yield in our experiment, its connection to the experimentally varied intensities of the SPP field and the probe laser is not directly evident. In the following we will map the fourth-order expectation values in the transition rate to products of the intensities of the fields. We first notice that not all expectation values in the transition rate are independent. In fact, if we consider the canoncial commutation relations for the creation and annihilation operators for the probe field and the SPP

field, we find that

$$\langle \hat{n}_{\text{Probe}}(\hat{n}_{\text{Probe}} - 1)\rangle_{\hat{\sigma}} = \langle \hat{a}^\dagger \hat{a}^\dagger \hat{a} \hat{a}\rangle_{\hat{\sigma}} \tag{S52a}$$

$$\langle \hat{n}_{\text{SPP}}(\hat{n}_{\text{SPP}} - 1)\rangle_{\hat{\sigma}} = \langle \hat{b}^\dagger \hat{b}^\dagger \hat{b} \hat{b}\rangle_{\hat{\sigma}} \tag{S52b}$$

$$\langle \hat{n}_{\text{Probe}} \hat{n}_{\text{SPP}}\rangle_{\hat{\sigma}} = \langle \hat{a}^\dagger \hat{b}^\dagger \hat{a} \hat{b}\rangle_{\hat{\sigma}} = \langle \hat{b}^\dagger \hat{a}^\dagger \hat{b} \hat{a}\rangle_{\hat{\sigma}} = \langle \hat{b}^\dagger \hat{a}^\dagger \hat{a} \hat{b}\rangle_{\hat{\sigma}} = \langle \hat{a}^\dagger \hat{b}^\dagger \hat{b} \hat{a}\rangle_{\hat{\sigma}}, \tag{S52c}$$

$$\langle \hat{c}\, \hat{n}_{\text{Probe}}\rangle_{\hat{\sigma}} = \langle \hat{a}^\dagger \hat{a}^\dagger \hat{a} \hat{b}\rangle_{\hat{\sigma}} = \langle \hat{a}^\dagger \hat{a}^\dagger \hat{b} \hat{a}\rangle_{\hat{\sigma}}, \tag{S52d}$$

$$\langle \hat{n}_{\text{SPP}} \hat{c}\rangle_{\hat{\sigma}} = \langle \hat{a}^\dagger \hat{b}^\dagger \hat{b} \hat{b}\rangle_{\hat{\sigma}} = \langle \hat{b}^\dagger \hat{a}^\dagger \hat{b} \hat{b}\rangle_{\hat{\sigma}}, \tag{S52e}$$

$$\langle \hat{n}_{\text{Probe}} \hat{c}^\dagger\rangle_{\hat{\sigma}} = \langle \hat{b}^\dagger \hat{a}^\dagger \hat{a} \hat{a}\rangle_{\hat{\sigma}} = \langle \hat{a}^\dagger \hat{b}^\dagger \hat{a} \hat{a}\rangle_{\hat{\sigma}}, \tag{S52f}$$

$$\langle \hat{c}^\dagger \hat{n}_{\text{SPP}}\rangle_{\hat{\sigma}} = \langle \hat{b}^\dagger \hat{b}^\dagger \hat{b} \hat{a}\rangle_{\hat{\sigma}} = \langle \hat{b}^\dagger \hat{b}^\dagger \hat{a} \hat{b}\rangle_{\hat{\sigma}}, \tag{S52g}$$

$$\langle \hat{c}\hat{c}\rangle = \langle \hat{a}^\dagger \hat{a}^\dagger \hat{b} \hat{b}\rangle_{\hat{\sigma}}, \tag{S52h}$$

$$\left\langle \hat{c}^\dagger \hat{c}^\dagger \right\rangle = \langle \hat{b}^\dagger \hat{b}^\dagger \hat{a} \hat{a}\rangle_{\hat{\sigma}}, \tag{S52i}$$

where we introduced the occupation number operators for the probe field $\hat{n}_{\text{Probe}} = \hat{a}^\dagger \hat{a}$ and the SPP field $\hat{n}_{\text{SPP}} = \hat{b}^\dagger \hat{b}$, and the mutual coherences $\hat{c} = \hat{a}^\dagger \hat{b}$ and $\hat{c}^\dagger = \hat{b}^\dagger \hat{a}$. In other words, all fourth-order expectation values in the field correlation function can be written as expectation values of second-order products of the occupation number operators and the mutual coherences.

To deduce the scaling of the different terms in Eq. (S51) as a function of the experimentally varied intensities, it is instructive to discuss Eq. (S52) under the assumption that the probe field and the SPP field initially are in a coherent product state $|\alpha\rangle \otimes |\beta\rangle$. Since the coherent states $|\alpha\rangle$ and $|\beta\rangle$ of the probe and SPP field satisfy the elementary relations

$$\hat{a}|\alpha\rangle = \alpha|\alpha\rangle = e^{i\phi_\alpha}\sqrt{\langle \hat{n}_{\text{Probe}}\rangle_{\hat{\sigma}}}|\alpha\rangle \tag{S53a}$$

$$\text{and} \quad \hat{b}|\beta\rangle = \beta|\beta\rangle = e^{i\phi_\beta}\sqrt{\langle \hat{n}_{\text{SPP}}\rangle_{\hat{\sigma}}}|\beta\rangle, \tag{S53b}$$

we can easily factorize every normally-ordered product of creation and annihilation operators in Eq. (S52) into simple products of complex numbers. We thus arrive at

$$\langle \hat{n}_{\text{Probe}}(\hat{n}_{\text{Probe}} - 1)\rangle_{\hat{\sigma}} = \langle \hat{n}_{\text{Probe}}\rangle^2_{\hat{\sigma}}, \tag{S54a}$$

$$\langle \hat{n}_{\text{SPP}}(\hat{n}_{\text{SPP}} - 1)\rangle_{\hat{\sigma}} = \langle \hat{n}_{\text{SPP}}\rangle^2_{\hat{\sigma}}, \tag{S54b}$$

$$\langle \hat{n}_{\text{Probe}} \hat{n}_{\text{SPP}}\rangle_{\hat{\sigma}} = \langle \hat{n}_{\text{Probe}}\rangle_{\hat{\sigma}} \langle \hat{n}_{\text{SPP}}\rangle_{\hat{\sigma}}, \tag{S54c}$$

$$\langle \hat{c}\, \hat{n}_{\text{Probe}}\rangle_{\hat{\sigma}} = e^{i\phi} \sqrt{\langle \hat{n}_{\text{Probe}}\rangle_{\hat{\sigma}}}^3 \sqrt{\langle \hat{n}_{\text{SPP}}\rangle_{\hat{\sigma}}}, \tag{S54d}$$

$$\langle \hat{n}_{\text{SPP}} \hat{c}\rangle_{\hat{\sigma}} = e^{i\phi} \sqrt{\langle \hat{n}_{\text{Probe}}\rangle_{\hat{\sigma}}} \sqrt{\langle \hat{n}_{\text{SPP}}\rangle_{\hat{\sigma}}}^3, \tag{S54e}$$

$$\langle \hat{n}_{\text{Probe}} \hat{c}^\dagger\rangle_{\hat{\sigma}} = e^{-i\phi} \sqrt{\langle \hat{n}_{\text{Probe}}\rangle_{\hat{\sigma}}}^3 \sqrt{\langle \hat{n}_{\text{SPP}}\rangle_{\hat{\sigma}}}, \tag{S54f}$$

$$\langle \hat{c}^\dagger \hat{n}_{\text{SPP}}\rangle_{\hat{\sigma}} = e^{-i\phi} \sqrt{\langle \hat{n}_{\text{Probe}}\rangle_{\hat{\sigma}}} \sqrt{\langle \hat{n}_{\text{SPP}}\rangle_{\hat{\sigma}}}^3, \tag{S54g}$$

$$\langle \hat{c}\hat{c}\rangle = e^{2i\phi} \langle \hat{n}_{\text{Probe}}\rangle_{\hat{\sigma}} \langle \hat{n}_{\text{SPP}}\rangle_{\hat{\sigma}}, \tag{S54h}$$

$$\left\langle \hat{c}^\dagger \hat{c}^\dagger \right\rangle = e^{-2i\phi} \langle \hat{n}_{\text{Probe}}\rangle_{\hat{\sigma}} \langle \hat{n}_{\text{SPP}}\rangle_{\hat{\sigma}}, \tag{S54i}$$

with the phase difference of the coherent states $\phi = \phi_\beta - \phi_\alpha$. We can now combine Eq. (S52) and Eq. (S54) and insert them into the transition rate Eq. (S51) to arrive at

$$\begin{aligned}
\Gamma_{\text{2PPE}}^{\text{quantum}}(\mathbf{K}, \Delta t) &= \frac{\mathcal{M}}{9\hbar^4}(2\pi)^2 \delta(\mathbf{K}) \left(\frac{\hbar\omega}{2\varepsilon_0 V}\right)^2 \left(\frac{\langle \hat{n}_{\text{Probe}} \rangle_{\hat{\sigma}}^2}{\mathcal{N}_{\text{Probe}}^2} + \left(1 + |\boldsymbol{\epsilon}_{\text{SPP}}^\perp|^2\right)^2 \frac{\langle \hat{n}_{\text{SPP}} \rangle_{\hat{\sigma}}^2}{\mathcal{N}_{\text{SPP}}^2}\right) \\
&+ 2\frac{\mathcal{M}}{9\hbar^4}(2\pi)^2 \delta(\mathbf{K}) \left(\frac{\hbar\omega}{2\varepsilon_0 V}\right)^2 \frac{\langle \hat{n}_{\text{Probe}} \rangle_{\hat{\sigma}} \langle \hat{n}_{\text{SPP}} \rangle_{\hat{\sigma}}}{\mathcal{N}_{\text{Probe}} \mathcal{N}_{\text{SPP}}} \left(2 + |\boldsymbol{\epsilon}_{\text{SPP}}^\perp|^2\right) \\
&+ 2\frac{\mathcal{M}}{9\hbar^4}(2\pi)^2 \left(\frac{\hbar\omega}{2\varepsilon_0 V}\right)^2 \frac{\sqrt{\langle \hat{n}_{\text{Probe}} \rangle_{\hat{\sigma}}} \sqrt{\langle \hat{n}_{\text{SPP}} \rangle_{\hat{\sigma}}}}{\sqrt{\mathcal{N}_{\text{Probe}}} \sqrt{\mathcal{N}_{\text{SPP}}}} \\
&\quad \times \left(\frac{\langle \hat{n}_{\text{Probe}} \rangle_{\hat{\sigma}}}{\mathcal{N}_{\text{Probe}}} + \left(1 + |\boldsymbol{\epsilon}_{\text{SPP}}^\perp|^2\right) \frac{\langle \hat{n}_{\text{SPP}} \rangle_{\hat{\sigma}}}{\mathcal{N}_{\text{SPP}}}\right) \\
&\quad \times \left(\delta(\mathbf{K} - \mathbf{k}_{\text{SPP}})e^{-i(\omega \Delta t - \phi)} + \delta(\mathbf{K} + \mathbf{k}_{\text{SPP}})e^{i(\omega \Delta t - \phi)}\right) \\
&+ \frac{\mathcal{M}}{9\hbar^4}(2\pi)^2 \left(\frac{\hbar\omega}{2\varepsilon_0 V}\right)^2 \frac{\langle \hat{n}_{\text{Probe}} \rangle_{\hat{\sigma}} \langle \hat{n}_{\text{SPP}} \rangle_{\hat{\sigma}}}{\mathcal{N}_{\text{Probe}} \mathcal{N}_{\text{SPP}}} \\
&\quad \times \left(\delta(\mathbf{K} - 2\mathbf{k}_{\text{SPP}})e^{-2i(\omega \Delta t - \phi)} + \delta(\mathbf{K} + 2\mathbf{k}_{\text{SPP}})e^{2i(\omega \Delta t - \phi)}\right).
\end{aligned} \quad (\text{S55})$$

Finally, we substitute the expectation values of the occupation numbers by the respective expectation values of the intensities, that is use the identities

$$\langle \hat{I}_{\text{Probe}} \rangle_{\hat{\sigma}} = \frac{1}{2} c_{\text{Probe}} \varepsilon_0 \langle \hat{\mathbf{E}}_{\text{Probe}}^{(-)} \cdot \hat{\mathbf{E}}_{\text{Probe}}^{(+)} \rangle_{\hat{\sigma}} = \frac{1}{2} c_{\text{Probe}} \varepsilon_0 \frac{\hbar\omega}{2\varepsilon_0 V} \frac{1}{\mathcal{N}_{\text{Probe}}} \langle \hat{n}_{\text{Probe}} \rangle_{\hat{\sigma}} \quad (\text{S56a})$$

$$\langle \hat{I}_{\text{SPP}} \rangle_{\hat{\sigma}} = \frac{1}{2} c_{\text{SPP}} \varepsilon_0 \langle \hat{\mathbf{E}}_{\text{SPP}}^{(-)} \cdot \hat{\mathbf{E}}_{\text{SPP}}^{(+)} \rangle_{\hat{\sigma}} = \frac{1}{2} c_{\text{SPP}} \varepsilon_0 \frac{\hbar\omega}{2\varepsilon_0 V} \frac{1 + |\boldsymbol{\epsilon}_{\text{SPP}}^\perp|^2}{\mathcal{N}_{\text{SPP}}} \langle \hat{n}_{\text{SPP}} \rangle_{\hat{\sigma}}, \quad (\text{S56b})$$

where $c_{\text{Probe}}$ and $c_{\text{SPP}}$ are the propagation speed of the probe field and the SPP field. With this we rewrite the transition rate in Eq. (S55) in terms of the intensities of the fields as

$$\begin{aligned}
\Gamma_{\text{2PPE}}^{\text{quantum}}(\mathbf{K}, \Delta t) &= \frac{\mathcal{M}}{9\hbar^4} \left(\frac{4\pi}{\varepsilon_0}\right)^2 \delta(\mathbf{K}) \left(\frac{\langle \hat{I}_{\text{Probe}} \rangle_{\hat{\sigma}}^2}{c_{\text{Probe}}^2} + \frac{\langle \hat{I}_{\text{SPP}} \rangle_{\hat{\sigma}}^2}{c_{\text{SPP}}^2}\right) \\
&+ 2\frac{\mathcal{M}}{9\hbar^4} \left(\frac{4\pi}{\varepsilon_0}\right)^2 \delta(\mathbf{K}) \frac{\langle \hat{I}_{\text{Probe}} \rangle_{\hat{\sigma}} \langle \hat{I}_{\text{SPP}} \rangle_{\hat{\sigma}}}{c_{\text{Probe}} c_{\text{SPP}}} \frac{2 + |\boldsymbol{\epsilon}_{\text{SPP}}^\perp|^2}{1 + |\boldsymbol{\epsilon}_{\text{SPP}}^\perp|^2} \\
&+ 2\frac{\mathcal{M}}{9\hbar^4} \left(\frac{4\pi}{\varepsilon_0}\right)^2 \sqrt{\frac{\langle \hat{I}_{\text{Probe}} \rangle_{\hat{\sigma}} \langle \hat{I}_{\text{SPP}} \rangle_{\hat{\sigma}}}{c_{\text{Probe}} c_{\text{SPP}} (1 + |\boldsymbol{\epsilon}_{\text{SPP}}^\perp|^2)}} \left(\frac{\langle \hat{I}_{\text{Probe}} \rangle_{\hat{\sigma}}}{c_{\text{Probe}}} + \frac{\langle \hat{I}_{\text{SPP}} \rangle_{\hat{\sigma}}}{c_{\text{SPP}}}\right) \\
&\quad \times \left(\delta(\mathbf{K} - \mathbf{k}_{\text{SPP}})e^{-i(\omega \Delta t - \phi)} + \delta(\mathbf{K} + \mathbf{k}_{\text{SPP}})e^{i(\omega \Delta t - \phi)}\right) \\
&+ \frac{\mathcal{M}}{9\hbar^4} \left(\frac{4\pi}{\varepsilon_0}\right)^2 \frac{\langle \hat{I}_{\text{Probe}} \rangle_{\hat{\sigma}} \langle \hat{I}_{\text{SPP}} \rangle_{\hat{\sigma}}}{c_{\text{Probe}} c_{\text{SPP}} (1 + |\boldsymbol{\epsilon}_{\text{SPP}}^\perp|^2)} \\
&\quad \times \left(\delta(\mathbf{K} - 2\mathbf{k}_{\text{SPP}})e^{-2i(\omega \Delta t - \phi)} + \delta(\mathbf{K} + 2\mathbf{k}_{\text{SPP}})e^{2i(\omega \Delta t - \phi)}\right).
\end{aligned} \quad (\text{S57})$$

This result is a direct quantum-mechanical analogue of Eq. (S14). It should again be highlighted, that the coherent state approximation for the transition rate as a function of the intensities is an approximation and should only exemplify how to map the observed intensity scalings of the different measurement signatures to the different fourth-order correlation functions between the SPP and the probe field. Ultimately, Eq. (S51) is more appropriate to understand the microscopic quantum-mechanical processes giving rise to the observed measurement signatures.

It is, however, possible to incorporate deviations from the idealized coherent states of the SPP and probe field into Eq. (S57). For example, instead of assuming that the probe and the SPP field are in coherent states one could factor out in Eq. (S51) the commonly used second-order (mutual) coherence functions

$$g^{(2)}_{\text{Probe,Probe}} = \frac{\langle \hat{n}_{\text{Probe}}(\hat{n}_{\text{Probe}} - 1)\rangle_{\hat{\sigma}}}{\langle \hat{n}_{\text{Probe}}\rangle^2_{\hat{\sigma}}} \tag{S58a}$$

$$g^{(2)}_{\text{SPP,SPP}} = \frac{\langle \hat{n}_{\text{SPP}}(\hat{n}_{\text{SPP}} - 1)\rangle_{\hat{\sigma}}}{\langle \hat{n}_{\text{SPP}}\rangle^2_{\hat{\sigma}}} \tag{S58b}$$

$$g^{(2)}_{\text{Probe,SPP}} = \frac{\langle \hat{n}_{\text{Probe}}\hat{n}_{\text{SPP}}\rangle_{\hat{\sigma}}}{\langle \hat{n}_{\text{Probe}}\rangle_{\hat{\sigma}} \langle \hat{n}_{\text{SPP}}\rangle_{\hat{\sigma}}} \tag{S58c}$$

$$g^{(2)}_{c,\text{Probe}} = \frac{\langle \hat{c}\,\hat{n}_{\text{Probe}}\rangle_{\hat{\sigma}}}{\sqrt{\langle \hat{n}_{\text{Probe}}\rangle^3_{\hat{\sigma}}}\sqrt{\langle \hat{n}_{\text{SPP}}\rangle_{\hat{\sigma}}}} \tag{S58d}$$

$$g^{(2)}_{\text{SPP},c} = \frac{\langle \hat{n}_{\text{SPP}}\hat{c}\rangle_{\hat{\sigma}}}{\sqrt{\langle \hat{n}_{\text{SPP}}\rangle^3_{\hat{\sigma}}}\sqrt{\langle \hat{n}_{\text{Probe}}\rangle_{\hat{\sigma}}}} \tag{S58e}$$

$$g^{(2)}_{c,c} = \frac{\langle \hat{c}\hat{c}\rangle_{\hat{\sigma}}}{\langle \hat{n}_{\text{Probe}}\rangle_{\hat{\sigma}} \langle \hat{n}_{\text{SPP}}\rangle_{\hat{\sigma}}}, \tag{S58f}$$

and then again follow our derivation. One would then arrive at a generalization of Eq. (S57), where each contribution is weighted by the respective second-order coherence function. Such a formulation of our theory might enable to deduce deviations from the idealized coherent state picture directly from experimental data. For example, if instead the SPP and the probe field would initially be in a Fock product state all terms involving the coherence $\hat{c}$ would vanish and thus no interference would be visible in the experiment.

# Supplementary Note 2: Background subtraction

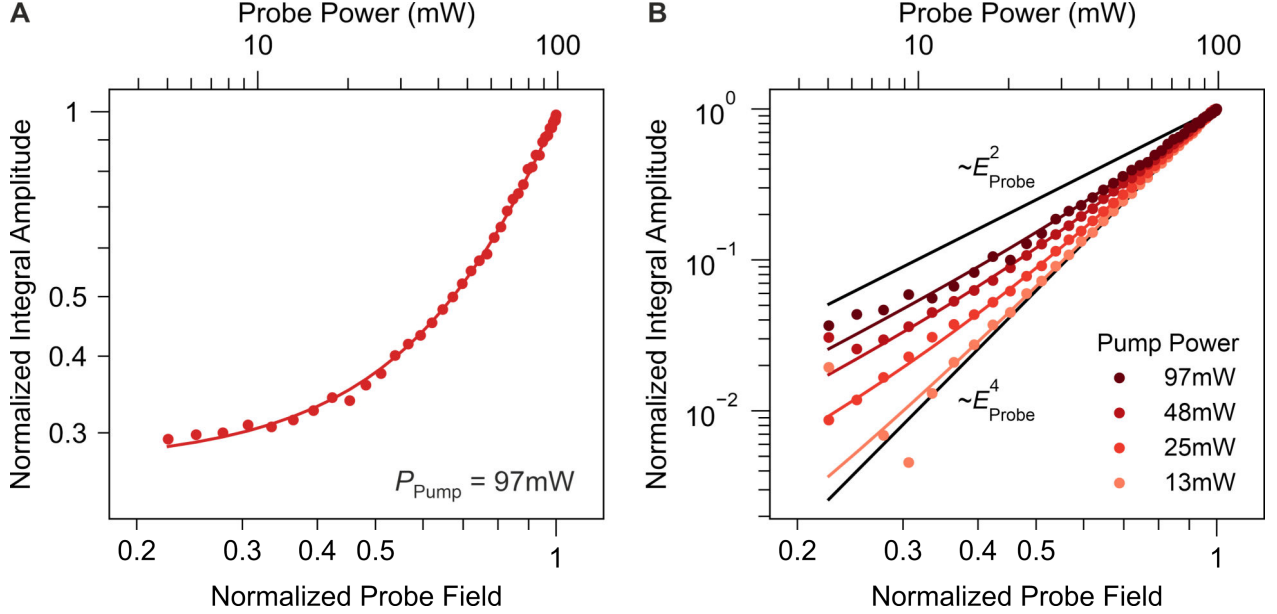

Figure S1: (a) Exemplary data for the electron yield at $\mathbf{K} = 0$ as a function of the probe power. The polynomial fit to the data is show as the red line. (b) Background corrected data obtained by fitting the electron yield for each pump power and subtracting the fit constant $c_0$.

In the main text we study the dependence of the electron yield in reciprocal space on the power of the probe laser pulses. Our theoretical treatment in the previous section shows, that the component of the electron yield with $\mathbf{K} = 0$ consists of contributions proportional to $I_{\text{Probe}}^2$, $I_{\text{SPP}}^2$, and $I_{\text{Probe}}I_{\text{SPP}}$. Additionally, in the experiment there is also a contribution to the electron yield at $\mathbf{K} = 0$, which is due to two-photon absorption directly from the probe laser pulses. Figure S1a exemplarily shows the electron yield at $\mathbf{K} = 0$ as a function of the probe power on a double-logarithmic scale. It is evident, that the different contributions to the electron yield can not directly be distinguished from the data because of the constant background contributions that do not depend on the probe power. We thus fit the data by a fourth-order polynomial

$$Y(\mathbf{K} = 0) = c_4 E_{\text{Probe}}^4 + c_2 E_{\text{Probe}}^2 + c_0, \tag{S59}$$

which is shown by the red line in Fig. S1a. The coefficients $c_i$ are the fitting parameters, where the constant background $c_0$ includes the contributions by plasmoemission proportional to $I_{\text{SPP}}^2 = E_{\text{SPP}}^4$ and by direct two-photon absorption from the pump laser pulses. The fitting constants $c_2$ and $c_0$, however, depend on the pump power, since both the SPP field-strength and the two-photon absorption from the pump are both pump power dependent. We can correct for the background contribution by subtracting the fitted value of $c_0$ for each pump power from the experimental data and arrive at the data in Fig. S1b, which is discussed in the main text.

# Supplementary References